\documentclass[sigconf]{acmart}

\usepackage{stfloats}
\usepackage[utf8]{inputenc}
\usepackage{adjustbox}
\usepackage{subcaption}
\usepackage{tabularx}
\usepackage{booktabs} 
\usepackage[normalem]{ulem}
\usepackage{float}
\usepackage{wrapfig}
\usepackage{etoolbox}
\usepackage{newunicodechar}
\newunicodechar{↔}{\leftrightarrow}
\usepackage{hyperref}
\usepackage[utf8]{inputenc}
\usepackage{multirow}
\usepackage{amsmath}
\usepackage{pgfplots}
\pgfplotsset{compat=1.18}

\newcommand{\mm}[1]{\textcolor{black}{#1}}

\AtBeginDocument{%
  }

\newcommand{\SEView}{\textsc{Shared Embodied View}}
\newcommand{\OOBView}{\textsc{Out-of-body View}}
\newcommand{\EAView}{\textsc{Embedded Anchored View}}

\copyrightyear{2026}
\acmYear{2026}
\setcopyright{cc}
\setcctype{by}
\acmConference[CHI '26]{Proceedings of the 2026 CHI Conference on Human Factors in Computing Systems}{April 13--17, 2026}{Barcelona, Spain}
\acmBooktitle{Proceedings of the 2026 CHI Conference on Human Factors in Computing Systems (CHI '26), April 13--17, 2026, Barcelona, Spain}
\acmPrice{}
\acmDOI{10.1145/3772318.3791433}
\acmISBN{979-8-4007-2278-3/2026/04}

\begin{document}


\begin{teaserfigure}
    \centering
     \includegraphics[width=0.98\textwidth]{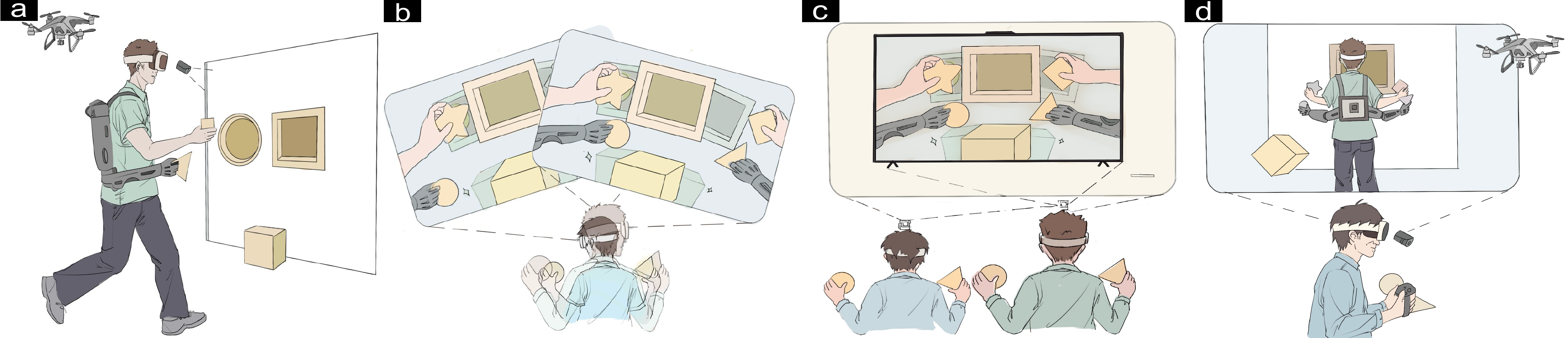} 
    \caption{Illustration of three perspectives for remote collaborative teleoperation using virtual supernumerary limbs (VSLs): (a) \emph{Host} user physically walking with VSLs attached to their back; (b) \SEView{}: the \emph{guest} shares the \emph{host}'s position with independent head rotation; (c) \EAView{}: the \emph{guest} views from the \emph{host}'s head location through a stabilized virtual window; (d) \OOBView{}: the \emph{guest} independently navigates a drone-like external viewpoint using a handheld controller \mm{(the drone serves as a visual metaphor for the decoupled virtual camera).}}
    \label{conceptimplementation}
    \Description{The figure shows four panels (a–d) representing three viewpoint modes for collaborative teleoperation using virtual supernumerary limbs (VSLs). Panel (a) depicts the \emph{host} user walking in VR with two robotic limbs mounted on their back. Panel (b) illustrates the \SEView{}, where the \emph{guest} shares the host’s position and body orientation but can independently rotate their head to look around. Panel (c) shows the \EAView{}, where the guest views the scene from the host’s head location through a stabilized virtual window, helping reduce visual discomfort. Panel (d) shows the \OOBView{}, where the guest uses a handheld joystick to control a drone-like external camera, allowing free movement for enhanced spatial awareness. Together, these modes allow dynamic viewpoint switching and asymmetric collaboration in VR.}
\end{teaserfigure}

\title[One Body, Two Minds]{One Body, Two Minds: Alternating VR Perspective During Remote Teleoperation of Supernumerary Limbs}



\author{Hongyu Zhou}
\orcid{0009-0007-3278-2122}
\affiliation{%
  \institution{The University of Sydney}
  \department{School of Computer Science}
  \country{Australia}
}
\email{hzho4130@uni.sydney.edu.au}

\author{Xincheng Huang}
\orcid{0000-0001-6923-6490}
\affiliation{%
  \institution{University of British Columbia}
  \department{Department of Computer Science}
  \country{Canada}
}
\email{xincheng.huang@ubc.ca}

\author{Winston Wijaya}
\orcid{0009-0003-8605-0681}
\affiliation{%
  \institution{The University of Sydney}
  \department{School of Computer Science}
  \country{Australia}
}
\email{wwij0922@uni.sydney.edu.au}

\author{Yi Fei Cheng}
\orcid{0000-0002-6027-4236}
\affiliation{%
  \institution{Carnegie Mellon University}
  \department{Human-Computer Interaction Institute}
  \city{Pittsburgh, Pennsylvania}
  \country{USA}
}
\email{yifeic2@andrew.cmu.edu}

\author{David Lindlbauer}
\orcid{0000-0002-0809-9696}
\affiliation{%
  \institution{Carnegie Mellon University}
  \department{Human-Computer Interaction Institute}
  \city{Pittsburgh, Pennsylvania}
  \country{USA}
}
\email{davidlindlbauer@cmu.edu}

\author{Eduardo Velloso}
\orcid{0000-0003-4414-2249}
\affiliation{%
  \institution{The University of Sydney}
  \department{School of Computer Science}
  \country{Australia}
}
\email{eduardo.velloso@sydney.edu.au}

\author{Andrea Bianchi}
\orcid{0000-0002-7500-7974}
\affiliation{%
  \institution{KAIST}
  \department{Department of Industrial Design}
  \city{Daejeon}
  \country{Republic of Korea}
}
\email{andrea@kaist.ac.kr}

\author{Zhanna Sarsenbayeva}
\orcid{0000-0002-1247-6036}
\affiliation{%
  \institution{The University of Sydney}
  \department{School of Computer Science}
  \country{Australia}
}
\email{zhanna.sarsenbayeva@sydney.edu.au}

\author{Anusha Withana}
\orcid{0000-0001-6587-1278}
\affiliation{%
  \institution{The University of Sydney}
  \department{School of Computer Science, Sydney Nano Institute}
  \country{Australia}
}
\email{anusha.withana@sydney.edu.au}

\renewcommand{\shortauthors}{Zhou et al.}

\begin{abstract}
Remote VR teleoperation with supernumerary robotic limbs enables distant users to operate in another’s local space. While a shared first-person view aids hand–eye coordination, locking the guest's camera to the host's head can degrade comfort, embodiment, and coordination. \mm{Based on a formative study (N=10) using a virtual supernumerary robotic limbs configuration to stress-test coordination, we propose guest-driven perspective switching from a shared first-person baseline (\SEView{}) to two alternatives: (a) a stabilized view with guest-controlled rotation (\EAView{}), and (b) a fully decoupled third-person view (\OOBView{}).} We ran a user study with 24 pairs (N=48), who switched between the baseline and proposed views as task demands changed. We measured performance, embodiment, fatigue, physiological arousal, and switching behaviors. Our results reveal role-dependent trade-offs: \OOBView{} improves navigation efficiency and reduces errors, while \EAView{} supports embodiment. We conclude with guidelines: use \EAView{} for hand-centric adjustments, \OOBView{} for navigation and object placement, and ensure smooth transitions.


\end{abstract}



\maketitle


\section{INTRODUCTION} 
\textit{``When you drift with someone, you feel like there’s nothing to talk about.''} This evocative quote from \textit{Pacific Rim~\cite{pacificrim2013}} describes the deep mutual understanding achieved through the fictional mental process that allows two users to co-pilot a single machine by sharing bodily sensations, intentions, and thoughts. It captures the promise of shared bodily control and sensorimotor synchrony. \mm{Though fictional, such scenarios are becoming increasingly plausible thanks to advances in telepresence and virtual reality (VR), enabling two individuals to act through a shared body, whether physical or virtual, in tightly coordinated ways~\cite{saraiji2018fusion,zhou2024coplayingvr,zhou2024pairplayvr}. }

Emerging robotic platforms, such as the humanoid Neo~\cite{1x_neo}, suggest a future where remote operators inhabit complex physical bodies. However, controlling locomotion while simultaneously performing precise manipulation can place high cognitive demands on a single operator~\cite{takizawa2019exploring}. Shared control is emerging in these scenarios, for instance, Multiple Operator, Single Robots (MOSR)\cite{Feth2009}. A natural emergence of this is co-embodiment, splitting control between a host and a guest, for instance, between navigation and manipulation. This, in turn, raises a central design question: \emph{viewpoint coupling}, that is, how tightly a guest user’s camera is linked to the host’s head and body motion, which directly affects how fluidly the pair can collaborate. In most VR systems, the remote first-person view is tied to the local user’s head; small head movements therefore propagate to the shared camera, destabilizing near-body control and blurring role boundaries~\cite{dass2024telemoma,zhu2023intention}. Simultaneously, advances in AI-based 3D scene reconstruction and world models are making it feasible for teleoperators to work inside navigable virtual replicas of real environments~\cite{kang2020review,kerbl20233d,marble_ai,peppa2020urban,GaussianNexus2025Huang}. In such digital twins, anchored or ``drone-like'' viewpoints can be realised virtually, even while the robot remains grounded in the physical world. Therefore, studying shared viewpoint control in VR offers a way to probe how co-embodied collaborators should manage perspectives for future telepresence systems. In this work, we focus on viewpoint management and shared visualisation, rather than simulating specific robot kinematics or control hardware.

\mm{In this paper, we developed a collaborative VR scenario in which two users, a \emph{host} and a \emph{guest}, share control over the same virtual body. To examine viewpoint management under demanding conditions, we augment this body with virtual supernumerary limbs (VSLs). We use this multi-limb configuration not as the sole target application, but as a challenging testbed that maximizes overlap in the near-body workspace, so that insights about viewpoint coupling can generalize to less complex shared-control scenarios.} A formative study with five participant pairs identified four empirical patterns: (1) prolonged visual coupling during locomotion co-occurring with \emph{guest} discomfort and blurred self-location; (2) limited perspective flexibility aligning with grasping errors; (3) VSL control instability during the \emph{host}'s head position changes; and (4) role ambiguity leading to \emph{host} pauses. Motivated by these observations, we adopt a task-contingent approach: the \emph{host} remains in a standard HMD first-person view, while the \emph{guest} may switch their viewpoint on demand to align coupling with navigation versus near-body phases.

Building upon these insights, we implemented three perspectives for the \emph{guest}: 
(i) \SEView{}, our design baseline, which co-locates the \emph{guest}'s camera with the \emph{host}’s head position. 
In this configuration, the \emph{guest}'s head position is locked to the \emph{host}, and moving the \emph{guest}'s head does not change the viewpoint location but only rotates the view direction independently. (ii) \EAView{} shows a stereoscopic portal near the \emph{host}, streaming a steady first-person view for the \emph{guest}; and (iii) \OOBView{}, a world-anchored, fully decoupled 6-DoF third-person camera for spatial awareness. 
Operationally, the \emph{host} remains in \SEView{} throughout. The \emph{guest} can switch between \SEView{} and either \EAView{} or \OOBView{}, and our study investigates two \emph{guest}-driven switching strategies: \SEView{}~$\leftrightarrow$~\EAView{} and \SEView{}~$\leftrightarrow$~\OOBView{}.

We then conducted a controlled, within-subjects study with 24 pairs of participants (N=48) to examine how perspective modes affect task performance, embodiment, workload/fatigue, physiology, and guest switching. Results show that \OOBView{} yielded fewer errors in the Factory task with comparable completion times, while participants switched to \EAView{} more frequently for precision-demanding subtasks. Guests reported higher subjective workload and fatigue in \OOBView{}, although HRV indicated lower physiological stress; participants tended to use \EAView{} for precision phases and \OOBView{} for navigation.

Hence, this research offers the following key contributions:

\begin{itemize}
\item \textbf{Empirical diagnosis.} A formative study diagnosed two systemic breakdowns, coordination friction and guest disorientation, arising from fixed first-person coupling in co-embodied multi-limb VR control.

\begin{sloppypar}
\item \textbf{System \& evaluation.} We introduce guest-driven perspective switching and evaluate two regimes
(\SEView{}\allowbreak \allowbreak$\leftrightarrow$\allowbreak \EAView{} \allowbreak\ \SEView{}\allowbreak \allowbreak$\leftrightarrow$\allowbreak \OOBView{})
in a controlled within\allowbreak-\allowbreak subjects study (N=48), measuring performance, workload\allowbreak\ /\allowbreak\ fatigue, embodiment, physiology, and switching behaviour.
\end{sloppypar}

\item \textbf{Design implications.} We derive guidelines on when to use \SEView{}, \EAView{}, and \OOBView{} to manage viewpoint coupling and support comfort, coordination, and embodiment in co-embodied multi-limb VR.

\end{itemize}

\section{RELATED WORK}
In this section, we contextualize our research within existing literature on perspective control, remote embodiment, and supernumerary robotic limbs. We highlight key insights and identify the knowledge gaps our study aims to address.

\subsection{Perspective Control}

Viewpoint selection significantly impacts remote collaboration and coordination, influencing both task performance and user experience. Here, we distinguish coordination from collaboration. Coordination stands for the alignment of actions and timing between collaborators, whereas collaboration refers to the broader process of jointly working toward a shared goal, including planning, decision-making, and mutual understanding~\cite{gronbaek2024blended}. Prior research has explored the effects of different perspectives on remote interactions and their impact on coordination. Galvan et al.~\cite{galvan2017characterizing} demonstrated how alternating between first-person and third-person perspectives can leverage the unique advantages of each viewpoint without significantly harming body ownership of the virtual body. Similarly, Komiyama et al.~\cite{komiyama2017jackin} constructed a system that allows users to freely switch between first-person and third-person views in a remote work setting, enabling precise task execution through first-person images while maintaining situational awareness via third-person viewpoints. Saraiji et al.~\cite{saraiji2018fusion} introduced a wearable system where a teleoperator controls two robotic limbs remotely, using an HMD to observe the environment from a shoulder-mounted perspective, highlighting the importance of viewpoint selection for skill-sharing and collaboration.

However, viewpoint discrepancies between collaborators remain a significant challenge, particularly in object-focused tasks where precise spatial alignment and gesture interpretation are crucial. Nagai et al.~\cite{nagai2015livesphere} explored 360-degree wearable cameras, allowing users to observe and communicate with an entire environment from a remote location, potentially enhancing collaborative spatial awareness. Similarly, Jones et al.~\cite{jones2015mechanics} highlighted the inherent limitations of handheld perspectives in mobile video chat, as collaborators struggled to perceive shared scenes clearly due to instability and narrow visual fields. Fussell et al.~\cite{fussell2003effects} found that scene-focused camera angles outperformed head-mounted perspectives in physical task coordination, as they provided better visibility of the shared workspace. Further, Tang et al.~\cite{tang2010three} showed that different perspectives optimize different task demands, with shared perspectives aiding textual reading, while asymmetric perspectives facilitate the creation of shared and personal workspaces.

\mm{Beyond egocentric and third-person views, prior work proposed techniques like Worlds-in-Miniature (WIM)~\cite{stoakley1995wim}, which offers a scaled-down, allocentric replica of the environment, later adapted for real-world control~\cite{seo2016hybrid}. Go-Go~\cite{poupyrev1996gogo} enables non-linear arm extension to reach distant objects. These methods expand the design space for remote perspective control and inform our investigation of dynamic viewpoint switching.} 

\mm{In summary, prior work shows that flexible viewpoint control can improve spatial awareness and collaboration in VR and telepresence. However, we still know little about how dynamic, user-driven perspective switching shapes coordination, embodiment, and fatigue when two users co-embody a single avatar and jointly control multiple limbs in a shared virtual environment.}
Our study introduces adaptive guest-driven switching between \SEView{}, \EAView{}, and \OOBView{}, to probe the trade-offs in control fluency, comfort, and visual stability under collaborative embodiment.

\subsection{Embodiment in Collaboration}
In this paper, we distinguish between ``self-embodiment'', which describes a user’s ownership/agency over their own avatar (commonly measured with AEQ~\cite{gonzalez2018avatar,zhou2025survey}), and ``remote/co-embodiment'' or ``social presence'', which describe mutual awareness of a partner. These constructs are related but distinct. In our evaluation, we quantify self-embodiment and user experience.

Self-embodiment relates to a user’s ownership, agency, and self-location with respect to their own avatar. Remote embodiment refers to visual proxies that convey a collaborator’s bodily state and support awareness and coordination. Our study quantifies self-embodiment while using lightweight partner cues to aid reference.
Research has explored methods where multiple users share control over a single avatar, using techniques like the weighted-average co-embodiment method, which distributes control percentages between participants~\cite{fribourg2020virtual,hagiwara2019shared,kodama2022enhancing}. Studies show that increasing control weight improves users' sense of agency, enhancing task coordination~\cite{kodama2022enhancing}. This method has been effective in contexts like VR-based rehabilitation, where even users with less control feel engaged~\cite{juan2023immersive,gonzalez2017immersive}.
Further work in multi-operator single robot (MOSR) systems examined how multiple users control a robot, later evolving to include shared control in VR environments with a focus on first-person perspectives and multisensory feedback, affecting psychological aspects like intention alignment~\cite{inami2022cyborgs,jeunet2018you,kennedy1993simulator}. The phenomenon of ``enfacement'', where users perceive a merged identity, highlights the immersive potential of shared control~\cite{schubert2001experience}.

Complementary to these control-centric approaches, effective collaboration also depends on how partners are represented and perceived, i.e., embodiment cues that support mutual awareness, reference, and non-verbal coordination. 
\mm{Partner embodiments span a spectrum from minimal surrogates such as the Telepointer~\cite{greenberg1996semantic}, 
through visualisations of partner hands~\cite{wong2015handson,zhou2024coplayingvr,tecchia20123d,Huang2024SurfShare,huang2024virtualnexus}, arms~\cite{doucette2013sometimes,tang2007videoarms,tang2010three,sodhi2013bethere}, 
and feet~\cite{alizadeh2016happyfeet}, to full-body AR/VR telepresence avatars~\cite{maimone2013general,orts2016holoportation,pejsa2016room2room}.} These representations improve joint attention, reference establishment, and social presence. When gestures alone are insufficient for object reference, raycasting and local scene reconstruction offer stable referents~\cite{duval2014improving,oda20123d}.

\mm{While remote embodiment is known to support collaboration, its interplay with dynamic perspective control remains underexplored in co-embodied VR scenarios. We examine this relationship in a co-embodied avatar, focusing on how shared-viewpoint schemes shape coordination and embodied experience.}

\subsection{Supernumerary Robotic Limbs}
Research on supernumerary robotic limbs (SRL) has investigated various ways of augmenting human bodily capabilities~\cite{prattichizzo2021human,saraiji2018metaarms,hu2017hand,vatsal2018design,Zhou2026SRLProxemics}, including additional robotic legs~\cite{parietti2015design}, fingers~\cite{hu2017hand}, and arms~\cite{vatsal2018design}. Our study focuses specifically on virtual supernumerary limbs (VSLs), aiming to expand user capabilities within virtual reality environments.

Initially, SRLs emerged from industrial contexts, designed to assist workers in demanding physical tasks such as construction and assembly operations~\cite{parietti2016supernumerary,parietti2013dynamic,parietti2014supernumerary}. Over time, the scope of SRL research broadened significantly, encompassing novel designs such as robotic tails to assist balance~\cite{maekawa2020dynamic,nabeshima2019arque}, and extra limbs facilitating complex manipulations~\cite{eden2022principles,prattichizzo2021human,tong2021review,yang2021supernumerary}. For instance, \citet{maekawa2020dynamic} presented a wearable robotic tail that enhanced stability during physically challenging tasks.
Control strategies for SRLs have consistently been a central research theme, with multiple intuitive mechanisms proposed to facilitate efficient limb manipulation. These include methods like remapping existing body movements (e.g., utilizing foot or shoulder gestures)\cite{saraiji2018metaarms,sasaki2017metalimbs,shimobayashi2021independent}, and employing brain-computer interfaces (BCI) for direct neural control\cite{penaloza2018towards}. While promising, many of these approaches depend heavily on fixed control mappings, which restrict adaptability to dynamic and varying task conditions. Further, comprehensive user evaluations remain scarce, as most studies are limited to concept validations involving single-user scenarios~\cite{saraiji2018metaarms,vatsal2017wearing,vatsal2018design}. In collaborative contexts, sharing body control between two users introduces additional demands, such as division of roles, viewpoint negotiation, and real-time coordination, that fixed mappings do not address.

In summary, SRLs and VSLs demonstrate substantial potential, while most prior evaluations emphasize single-user prototypes with fixed mappings. We complement prior work by studying guest-driven switching from \SEView{} to \EAView{} or \OOBView{}, improving coordination, efficiency, and comfort when two users co-control one body.

\section{FORMATIVE STUDY}
\label{sec:formative}

To surface early-stage challenges in shared-body collaboration during locomotion, we conducted a diagnostic formative study focused on the limitations of a fixed shared first-person view, which is commonly adopted in prior work~\cite{zhou2025juggling,saraiji2018fusion}. 

\subsection{Method}

\begin{figure}[htbp]
  \centering
  \includegraphics[width=\linewidth]{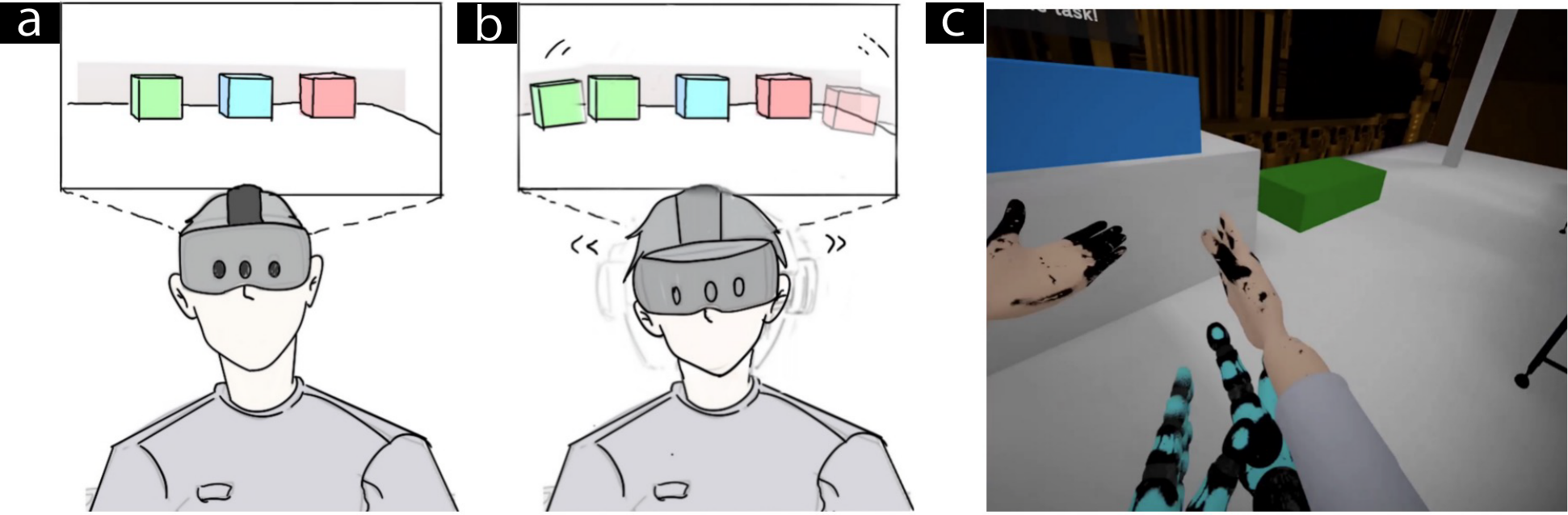}  
  \caption{Formative study configuration showing (a) the stable first-person perspective from the \emph{host}, (b) the \emph{guest}’s co-located but independently rotatable view illustrating visual instability, and (c) the avatar embodiment with VSLs used for near-body manipulation.}
  \label{Formative_study}
  \Description{figure comprising three panels: (a) shows the stable first-person perspective of the \emph{host} user, clearly depicting objects positioned linearly; (b) shows the independently controlled viewpoint of the \emph{guest} user with visual instability caused by head rotations, demonstrating challenges in visual comfort and object manipulation; (c) presents the actual VR environment, including detailed avatar limb positions and Virtual Supernumerary Limbs (VSLs), emphasizing the complexity of shared embodiment and joint manipulation tasks.}
\end{figure}

We recruited five dyads ($N=10$), each composed of a \emph{host} and a \emph{guest}, all with prior VR experience but no exposure to shared-control systems. This ensured that participants were not distracted by the novelty of VR itself or its basic controls, allowing us to better isolate usability challenges specific to shared-body interaction. Both participants were immersed in the same virtual environment via head-mounted displays.
The \emph{host} was embodied as a full-body avatar augmented with a pair of Virtual Supernumerary Limbs (VSLs) attached to their back. The \emph{guest} shared the \emph{host}’s first-person view via a camera rigidly anchored to the \emph{host}’s head position. While the \emph{guest} could not change the camera’s spatial location, they retained independent control over view orientation through head rotation. This configuration, which we term the \SEView, served as the baseline design condition (Figure~\ref{Formative_study}).

\noindent \textbf{\textit{Task and Procedure:}} Participants performed a simplified collaborative task designed to surface visual, operational, and embodiment challenges specific to shared-body teleoperation with VSLs. The task involved the joint transport of small boxes across a 2–3 meter space. The \emph{guest} used VSLs to grasp objects from a table; after a successful grasp, the \emph{host} navigated to a placement zone. \mm{Before the main trials, each pair had a short familiarisation period to practise grasping and transporting boxes using the VSLs. They then completed two 5-minute recorded trials. The short duration was intended to foreground early-stage coordination breakdowns and discomfort, rather than capture long-term adaptation effects.} We collected observational notes, head rotation logs, VSL activity traces, and think-aloud comments~\cite{hertzum2016usability,hertzum2015thinking}. Post-trial, participants were briefly interviewed on discomfort, control issues, and coordination experiences~\cite{dong2019multi}.

\noindent \textbf{\textit{Measures and Analysis.}}
To identify breakdowns in shared-body coordination under fixed first-person coupling, we used three diagnostic indicators: (i) prolonged head rotations by the \emph{host} (lasting $>$10\,s), which are known to induce discomfort in the \emph{guest}~\cite{stauffert2020latency,tian2022cybersickness}; (ii) grasping errors (misses or drops per attempt) to assess VSL precision~\cite{adkins2021grasping,oprea2019grasping}; and (iii) self-reported loss-of-control from the \emph{guest} and observed host pauses as markers of coordination breakdown~\cite{hertzum2015ta}. Two authors independently coded observation notes and interview/think-aloud transcripts using these three indicators as deductive anchors. Discrepancies were resolved through discussion, and related codes were grouped into macro-level themes following reflexive thematic analysis~\cite{braun2006using,braun2021one,vindrola2021rapid}. We triangulated qualitative excerpts with synchronized system logs (e.g., head rotation, VSL activity, locomotion events) to support interpretation~\cite{denzin1978researchact,carter2014triangulation}. Given the small sample size ($N=10$), we report dyad-level coverage, representative excerpts, and cross-modal log co-occurrence to inform design, not to claim generalizable causality~\cite{malterud2016information}.

\subsection{Findings}

\subsubsection{\noindent \textbf{Guest Control Drops Without Perspective Flexibility (5/5 dyads).}}
\emph{Guests} reported losing spatial orientation when the \emph{host} was moving. We define this as disorientation during \emph{host} movement that coincides with grasping errors. Participants remarked, ``I kept losing where the arms'' (\emph{guest}, P04) and ``felt like the hands lagged behind what I wanted to do'' (\emph{guest}, P08); \emph{hosts} noticed this too (``When I started walking, he said `wait','' \emph{host}, P03). Grasping errors often occurred shortly after the \emph{host} began moving, according to system logs—suggesting a close relationship, though not a causal one. These findings highlight the need for a viewpoint mode that maintains stable, body-centered alignment during fine motor tasks, allowing \emph{guests} to invoke it as needed for precise control.

\subsubsection{\noindent \textbf{Hosts Stop Moving Due to Role Confusion (3/5 dyads).}} 
Following prior work on detecting stillness in movement trajectories and HMD logs~\cite{bonavita2022stop,wagstaff2018lstmzvd}, we identified \emph{host} pauses as periods of minimal movement lasting several seconds during ongoing tasks. \emph{Hosts} often attributed these pauses to uncertainty about whether continued movement would interfere with the \emph{guest}'s manipulation. Participants remarked, ``I kept stopping, unsure if moving would disrupt them'' (\emph{host}, P03) and ``I waited here until he finished'' (\emph{host}, P05), with \emph{guests} corroborating this dynamic (``I needed to say `don’t move''' \emph{guest}, P06). These co-occurrences suggest hesitation rooted in unclear role boundaries, rather than deliberate coordination.

\begin{table*}
\centering
\small
\resizebox{\linewidth}{!}{
\begin{tabular}{lcccc}
\hline
\textbf{View} & \textbf{Camera Anchoring} & \textbf{Orientation} & \textbf{Translation} & \textbf{Render} \\
\hline
\SEView{} & Host-anchored & Guest head (independent) & No & Full-screen \\
\EAView{} & Decoupled-anchored & Guest head (linked) & No & Stabilized portal \\
\OOBView{} & World-anchored (free-floating) & Guest head / controller & Yes (6-DoF) & Full-screen \\
\hline
\end{tabular}
}
\caption{Differences across views along two axes: position anchoring and view control. Host always stays in \SEView{}; the guest switches between \SEView{} and one alternative per condition.}
\label{tab:view-comparison}
\end{table*}

\subsubsection{\noindent \textbf{Prolonged Shared View Causes Cybersickness (5/5 dyads).}}
When the \emph{guest} shared the host’s first-person view, extended walking or turning by the host often coincided with verbal reports of dizziness, disorientation, or difficulty controlling the arms. Participants described feeling dizzy or unsure of their location after several seconds of walking or turning, e.g., ``long turns made me dizzy'' (\emph{guest}, P02), and ``I [kept] losing where the arms [were], and felt nauseous'' (\emph{guest}, P06). Discomfort did not arise from sudden turns alone, but more often from sustained motion lasting over 10 seconds, even when the motion was low in intensity. In the system logs, these episodes overlapped with sustained head rotations or locomotion events. Grasping errors were also more likely to occur shortly after such movement sequences. Although we make no causal claims, this repeated co-occurrence points to a possible mismatch between visual motion and control stability during prolonged shared viewing.

\subsubsection{\noindent \textbf{Camera Shifts Disrupt Guest’s VSL Control (4/5 dyads).}}
When the \emph{guest} shared the host’s camera view, sudden or sharp head turns by the \emph{host} often disrupted the \emph{guest}’s ability to control the virtual arms. Participants described losing targeting accuracy or feeling disoriented during these transitions, such as ``Hard to keep control when everything shifts'' (\emph{guest}, P06) and ``every turn jolted my aim'' (\emph{guest}, P02). Triangulating with locomotion logs, these breakdowns frequently occurred within short windows after \emph{host} heading-change events, evidence of time-adjacent co-occurrence rather than causality.

\subsubsection{\noindent \textbf{Summary of Findings.}} Our formative study revealed recurring difficulties with a fixed shared first-person view. First, \emph{guests} often lost track of hand position during precise actions, especially when \emph{hosts} moved their head mid-task, leading to grasping errors and hesitation. At the same time, \emph{hosts} frequently paused movement, unsure whether the guest was in control, reflecting ambiguity in coordination roles. Second, shared locomotion episodes where the \emph{host} moved while the \emph{guest} remained visually coupled caused discomfort, disorientation, and disengagement for the guest. These issues clustered into two under-explored challenge areas: (C1) coordination breakdowns during precision tasks, and (C2) discomfort and control loss from shared movement. \mm{System logs confirmed that most errors clustered around host motion: about 71\% of logged errors (45/63) occurred within 4\,s of walking onset or a sharp head turn.} These findings highlight the need to restore guest perceptual agency and reduce coordination friction during task transitions.


\section{DESIGN AND IMPLEMENTATION}

To address the issues surfaced in our formative study, we designed two guest-driven perspective modes that directly target the observed challenges. In C1, coordination broke down during precise manipulation due to ambiguous visual control. To address this, we introduce the \EAView{}, which grants the guest independent rotational control of the viewpoint. This stabilizes the visual frame during fine-grained tasks and helps clarify interaction roles. In C2, guests experienced discomfort and disorientation caused by involuntary shared locomotion. To address this, we implemented the \OOBView{}, a third-person, freely positionable camera that decouples the guest’s perspective from the host’s body, easing spatial confusion and supporting higher-level navigation decisions. The specific distinctions between these perspective modes are summarized in Table~\ref{tab:view-comparison}.

\subsection{Embedded Anchored View (C1)}
We implemented \EAView{} as a body-anchored yet independently rotatable viewpoint, as shown in Figure~\ref{Out} (a). The system renders a stabilized egocentric feed as a stereoscopic portal, positioned at a fixed offset from the host and oriented toward the host’s head-camera direction. This provides visual stability while preserving directional control. We also apply light IPD tuning and frame smoothing to support accurate manipulation.

Previous research has demonstrated improved depth perception and task accuracy with slight inter-pupillary distance (IPD) adjustments in VR~\cite{guedry1998motion, nurnberger2021mismatch}. Thus, we slightly reduced the guest’s effective IPD within the EAV window to enhance binocular depth cues, facilitating precise grasping and alignment tasks. 

Moreover, as rapid \emph{host} movements can disrupt the \emph{guest}’s reference and degrade VSL stability~\cite{chen2018effect}, we employed Unity SmoothDamp stabilization (exponential smoothing) on EAV camera updates to soften \emph{host} jitter. This preserves a steady body-anchored reference during \emph{host} motion, improving comfort and control precision.

Finally, to reinforce role clarity, we emphasized visual embodiment cues known to reduce hesitation in collaboration~\cite{kornfield2021so}. While the shared avatar is visible in all modes, \EAView{} shows a steady 3D window that keeps the host’s hands, arms, and VSLs visually locked to the body with correct in-front/behind relationships, making the control locus near the body unambiguous.

\subsection{Out-of-body View (C2)}

\OOBView{} provides a fully decoupled, world-frame third-person camera with 6-DoF motion (rotation and translation), independent of the \emph{host}'s movement, as shown in Figure~\ref{Out} (b). The \emph{guest} controls the drone camera via joystick inputs. We apply smooth interpolation and bound speed/acceleration, with collision checks to prevent disruptive motion.

\begin{figure*}[htbp]
  \centering
  \includegraphics[width=\linewidth]{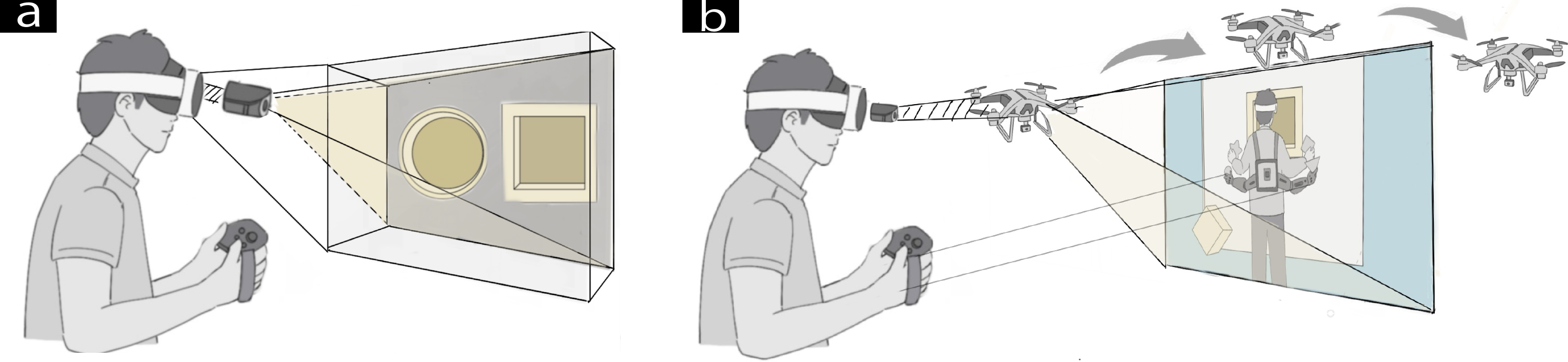}  
  \caption{Illustration of the \EAView{} and \OOBView{}. (a) The \emph{guest} user's viewpoint is spatially aligned with the \emph{host}'s viewpoint but rendered as a video-like window. (b) The \emph{guest} user independently controls a virtual drone camera with 6 degrees-of-freedom (6-DoF), enabling flexible positional and rotational adjustments independent of the \emph{host}'s movements \mm{(the drone serves as a visual metaphor for the decoupled virtual camera).}}
  \label{Out}
  \Description{(a) An illustration showing a remote (guest) user wearing a VR headset and using a controller, observing the collaborative workspace through a virtual window. A dedicated virtual camera rig aligned to the host's head position provides this independent view, allowing the guest user to rotate their viewpoint freely. The visual includes precise representations of spatial alignment and highlights improved stability through viewpoint smoothing, enhancing comfort and accuracy during tasks. (b) Illustration showing a remote (guest) user wearing a VR headset and operating a controller, independently controlling a drone-like virtual camera. The drone camera moves freely around a separate user (host) engaged in a VR task, who is depicted with Virtual Supernumerary Limbs (VSLs) performing simultaneous actions. The illustration visually represents the spatial independence of the guest user's viewpoint, highlighting smooth transitions between different camera positions to maintain visual comfort and stable environmental awareness.}
\end{figure*}

To improve the \emph{guest}’s spatial awareness and viewpoint flexibility, we implemented a 6-degree-of-freedom (6-DoF) virtual drone camera. The \emph{guest} controlled positional (forward, backward, lateral, vertical) and rotational (pitch, yaw, roll) movements using intuitive VR joystick inputs. Smooth camera transitions were ensured through spherical linear interpolation (Slerp) and linear interpolation (Lerp), enhancing visual comfort and spatial understanding.

Stabilizing VSL control during \emph{host}'s movements required a stable spatial reference frame. Thus, we implemented the drone camera in a fixed world coordinate system, independent of \emph{host} movement~\cite{kim2025effects}. Additionally, real-time collision detection and spatial constraints via Unity’s raycasting prevent visual disruptions and maintain stable associations with the shared environment and avatar~\cite{burg2022real}.

\subsection{Perspective Switching Mechanism}

In our system, \SEView{} is the default host-centric view and serves as the baseline from which other perspectives are invoked. To overcome limitations of this fixed shared view~\cite{bhandari2020influence,cao2018visually}, we implement a \emph{guest-driven} perspective-switching mechanism that lets users fluidly alternate between views based on task demands.

This mechanism allows the \emph{guest} to transition from \SEView{} to a condition-specific alternative view using VR controller input, which activates or deactivates dedicated Unity camera components. To preserve comfort and spatial context, each switch is rendered with short cross-fades, input debounce, and on-screen indicators. The design enables the \emph{guest} to time perspective transitions to task needs: invoking \EAView{}'s stabilization, body-aligned portal for fine manipulation, or \OOBView{}'s fully decoupled third-person view for spatial planning. This mechanism directly supports user agency, reduces prolonged visual coupling, and maintains stable VSL control during host movement—addressing the challenges identified in our formative study.

\subsection{VSLs Independent Control and Stabilization}

Formative findings indicated that VSL control became fragile under \emph{host} locomotion and heading changes. We therefore decouple VSL actuation from \emph{host} head/body transforms and drive VSL poses in a world-referenced frame initialized when the \emph{host} is stationary. Using Meta Quest 3 tracking, we monitor head orientation and translation, and flag short-window spikes in angular velocity or position delta as ``significant motion.'' When flagged, a stabilization mode engages: VSLs are either held in place or eased back to a predefined neutral pose via Unity \texttt{SmoothDamp} to avoid abrupt shifts. Guest input immediately disables stabilization and enters active control to preserve precision and responsiveness. This mechanism mitigates motion-induced disruptions without altering the \emph{host}'s baseline view and applies uniformly under \SEView{} and \EAView{} (the difference is rendering only).

\begin{figure}
\centering\includegraphics[width=1\linewidth]{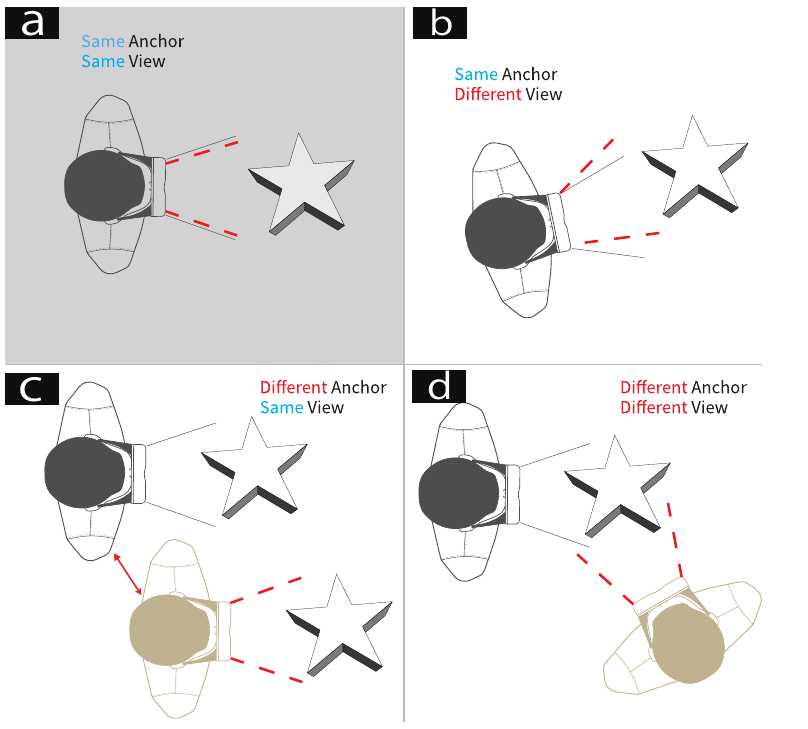}
\caption{\mm{Guest perspectives organized as an \textit{Anchor–View} matrix relative to the host. (a) \textbf{Locked Egocentric}: \emph{same anchor, same view}; guest camera is rigidly locked to the host’s head pose and view. (Shown for comparison; not studied, as it grants the guest no agency and renders them a passive observer under full host control). 
(b) \textbf{\SEView{}}: \emph{same anchor, different view}; guest shares the host’s position but rotates the camera independently. 
(c) \textbf{\EAView{}}: \emph{different anchor, same view}; guest remains independently positioned but views a stabilized window of the host’s egocentric feed. 
(d) \textbf{\OOBView{}}: \emph{different anchor, different view}; guest drives a free-floating 6-DoF drone camera, decoupled from the host.}}
\label{Out}
\Description{Top-down diagrams showing four guest-perspective modes relative to a host with back-mounted robotic arms. 
Panel (a): guest and host share both position and view; panel (b): guest shares host position but rotates view independently; 
panel (c): guest stays at a separate position but sees host’s egocentric feed in a stabilized window; 
panel (d): guest controls a free-flying 6-DoF camera with both position and view independent of the host.}
\end{figure}

\subsection{Implementation Details}
We developed the system in Unity 2022.3.7f1 and deployed it on two Meta Quest 3 headsets. Both users share one anthropomorphic avatar in VR: the \emph{host} navigates and controls the primary arms, and the \emph{guest} operates the back-mounted VSLs. 
Each headset was tethered via USB to a dedicated PC on the same local network, achieving latency below 5ms~\cite{fan2021high}. Multiplayer synchronization used Photon Unity Networking with a locally hosted Photon Server to minimize network hops. The \emph{host} and \emph{guest} were located in separate rooms to preserve a remote collaboration context while enabling natural voice communication via audio only.
Perspective control, VSL manipulation, and task logic were integrated into the Unity scene. On-screen indicators show the current mode. A logging module records head kinematics, perspective switches, VSL commands, and grasp events.


\section{EXPERIMENTAL METHOD}

We ran a within-subjects experiment to examine how two guest-controlled view-switching conditions shape user strategies and collaborative behaviors when operating VSLs with a shared-body avatar. Both tasks required tight coordination between the \emph{host} and \emph{guest}, emphasizing precision, timing, and role-dependent experience in dynamic VR scenarios.

To align with our revised research scope, we address the following RQs:

\begin{itemize}
    \item \textbf{RQ1}: How do \EAView{} and \OOBView{} affect subjective workload, fatigue, and physiological responses during collaboration?
    \item \textbf{RQ2}: How do \EAView{} and \OOBView{} influence self-embodiment (AEQ) for hosts and guests?
    \item \textbf{RQ3}: How do \EAView{} and \OOBView{} affect task performance (completion time, errors) in dynamic shared-body collaboration when switching is initiated from the \SEView{}?
    \item \textbf{RQ4}: When and why do \textit{guests} switch between \SEView{}\,$\leftrightarrow$\,\EAView{} or \SEView{}\,$\leftrightarrow$\,\OOBView{} as task demands evolve, and how are these switches associated with performance and experience?
\end{itemize}

\subsection{Experimental Conditions and Procedure}

\subsubsection{Experimental Conditions}

We tested two conditions: \SEView{}$\leftrightarrow$\EAView{} and \SEView{}$\leftrightarrow$\OOBView{}. In both conditions, tasks began in \SEView{}; the \emph{guest} could switch to the condition-specific alternative, while the \emph{host} remained in \SEView{} throughout. Each session comprised two task runs. Task type and condition order were fully counterbalanced across participant pairs to mitigate learning effects and fatigue, ensuring each pair experienced one run per condition.

\subsubsection{Participants}

We recruited 24 participant pairs (48 participants; 24 female, 24 male; M=22.8 yrs, SD=3.2 yrs; no color vision deficiencies). This sample size aligns with common practice in HCI research involving collaborative VR systems~\cite{caine2016local}. All participants had prior experience with VR~\cite{cummings2016immersive}. 16 participants reported regular VR use, while 32 primarily engaged with VR for entertainment. Each participant pair consisted of a \emph{host} and a \emph{guest}.

\subsection{Experimental Tasks}

\subsubsection{Transportation Task}

\begin{figure*}[htbp]
  \centering
  \includegraphics[width=\linewidth]{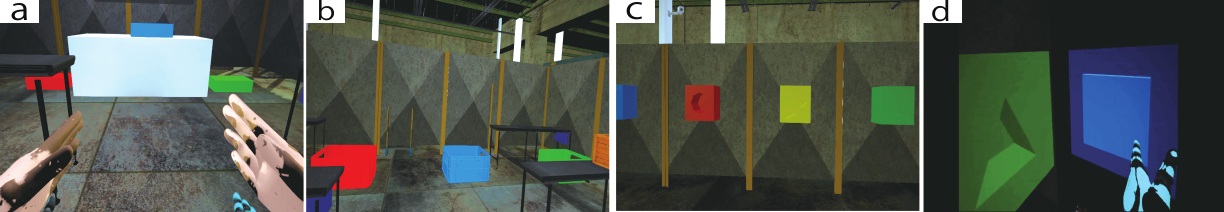}
  \caption{Task overview: Transportation and Factory. Panels (a–b) depict the \textit{Transportation} task: (a) colour-coded boxes placed on a central table; (b) matching colour-coded bins distributed throughout the VR environment. Panels \textbf{(c–d)} depict the \textit{Factory} task: (c) geometrically distinct holes on a wall panel as precise insertion targets; (d) collaborative insertion of objects (e.g., cubes) into the corresponding holes.}
  \label{fig:tasks_overview}
  \Description{A four-panel composite image. Panels (a) and (b) show the Transportation task: colour-coded boxes on a central table and matching bins placed around the room. Panels (c) and (d) show the Factory task: a wall panel with distinct target holes and the team inserting a matching geometric object into its hole.}
\end{figure*}

The first task simulated a mobile manipulation scenario commonly encountered in collaborative work and human-robot collaboration environments~\cite{wang2021shared,wongjirad2024development,de2021towards,ramasubramanian2021operator}, as shown in Figure~\ref{fig:tasks_overview} (a,b). In each trial, the participant pair (host and guest) started from a designated starting location. Their goal was to collaboratively transport colour-coded boxes placed centrally on a table into corresponding target bins distributed across a virtual room approximately 6$\times$8 meters in size, populated with static obstacles such as tables, walls, and barriers. The \emph{host} navigated the virtual environment freely by physically walking in the tracking space, ensuring balance, obstacle avoidance, and efficient positioning, while the \emph{guest} remotely controlled the VSLs using VR controllers, pressing specific buttons to grasp and release boxes accurately, a role distribution consistent with prior collaborative robotics studies~\cite{gallipoli2024virtual,yim2022wfh}. \mm{To enforce joint action, the boxes were designed to be large enough to require two hands to lift and carry, mimicking the size and weight of bulky real-world objects. While two hands stabilized the load, the remaining two were required to interact with the environment by pushing aside obstacles or lifting spring-loaded bin covers, creating situations in which all four hands had to operate simultaneously.}

This task required continuous collaboration: the \emph{host} needed to provide optimal positioning for the \emph{guest}'s manipulation tasks, whereas the \emph{guest} had to adapt to continuous perspective shifts caused by the \emph{host}'s movement. The design intentionally introduced continuous demands for spatial navigation and fine-grained manipulations, enabling systematic investigation of the impacts of dynamic viewpoint control, cognitive load, and collaborative efficiency in shared-body remote teleoperation scenarios.

\subsubsection{Factory Task}

The second task required participants to perform precision object manipulation and insertion tasks collaboratively, emphasizing precise spatial coordination and multi-limb synchronization~\cite{wu2023mr}, as shown in Figure~\ref{fig:tasks_overview}~(c,d). In each trial, participants simultaneously controlled the avatar’s four hands (two controlled by the \emph{host} and two by the \textit{guest}) to insert four geometrically distinct objects (e.g., cubes, cones, spheres, and cylinders) into corresponding matching holes on a virtual wall panel. Each insertion required precise alignment and steady positioning to succeed; insertion errors included misalignment, improper orientation, or unsuccessful attempts. The \emph{host} controlled the avatar's locomotion and two primary arms, while the \textit{guest} controlled the two extra arms. Clear visual feedback indicated successful insertion. \mm{To encourage multi-limb synchronization, we designed a timed batch-insertion task. All four target holes were always available, and participants were asked to complete all insertions before the trial time limit expired. The pick-up area and insertion panel were spatially separated, encouraging participants to carry multiple objects per trip when possible. Using all four hands in parallel helped reduce walking, minimize idle time between insertions, and improve overall efficiency. While not strictly required, coordinating all four hands became the most effective strategy for completing the task on time.} The task specifically aimed to evaluate how dynamic perspective-switching influenced precision tasks requiring fine motor coordination, collaborative synchronization, and role-specific embodiment under varying viewpoints.

\subsection{Experimental Data Collection}

To evaluate task performance and switching strategies, we collected quantitative data covering subjective embodiment, workload and fatigue, performance metrics, physiological responses, and qualitative data through interviews and think-aloud protocols. 

\subsubsection{Qualitative Data}

We recorded audio and video during the think-aloud session and post-task semi-structured interviews as qualitative data.

\subsubsection{Performance Measures}
Task performance metrics included completion time, error rate, and perspective-switching behaviors. In the Transportation Task, we recorded completion time (from grasp initiation to successful bin placement) and errors (failed grasps, drops, or wrong-bin placements). In the Factory Task, we measured insertion time per object and recorded insertion errors, which included both translational misalignments (incorrect positional offsets relative to the target hole) and rotational misalignments (incorrect orientation of objects relative to the required target orientation). Participants received no performance feedback during trials.

\subsubsection{Subjective Feedback}

We administered three standardized questionnaires:

The Avatar Embodiment Questionnaire (AEQ) measured self-embodiment, consistent with prior work on VR embodiment~\cite{gonzalez2018avatar}. Following prior SRL-in-VR work that administers AEQ in a pre/post-adaptation schedule \cite{arai2022embodiment}, we followed the same timing logic and extended it to our two perspective-switching strategies by collecting one pre-adaptation AEQ under \SEView{} and two post-adaptation AEQ: one after \SEView{}\allowbreak$\rightarrow$\allowbreak\,\EAView{} and one after \SEView{}\allowbreak$\rightarrow$\allowbreak\,\OOBView{}.

NASA Task Load Index (NASA-TLX) assessed perceived workload (mental, physical, temporal, effort, performance, frustration)~\cite{hart2006nasa}.
Visual Analog Scale for Fatigue (VAS-F) measured subjective fatigue across 18 dimensions~\cite{shahid2011visual}. \mm{While established measures of cybersickness such as the Simulator Sickness Questionnaire (SSQ)~\cite{kennedy1993simulator} are available, in this first study we focused on workload (NASA--TLX) and fatigue (VAS--F) as our primary subjective indicators of discomfort in shared viewpoints, to limit questionnaire burden and align the scales with our task demands. We acknowledge the absence of a dedicated cybersickness instrument as a limitation and see it as an important addition for future work.}

\subsubsection{Physiological Data - Heart Rate Monitoring}

We continuously recorded heart rate data using a Polar H10 chest strap~\cite{vermunicht2023validation}, known for accuracy and minimal interference with VR~\cite{kumpulainen2024assessing,rutkowski2021immersive}. Specifically, we analyzed heart rate variability (HRV) and peak heart rate during intensive task phases and critical perspective-switching events to quantify physiological correlates of subjective experience~\cite{alshanskaia2024heart,immanuel2023heart}. \mm{RR-interval series were preprocessed following psychophysiology recommendations~\cite{laborde2017heart}. We removed artefact-contaminated/ectopic intervals, interpolated only short gaps, and computed RMSSD in sliding windows during active task periods (full parameters in Appendix~\ref{app:physio-processing}).} Heart rate variability (HRV) was computed using the time-domain RMSSD index (root mean square of successive differences between adjacent NN intervals). Higher values indicate lower physiological stress.

\subsubsection{Procedure}

\mm{We adopted a fully within-subjects, repeated-measures design over two consecutive days. Each dyad experienced both collaborative tasks (Transportation and Factory) and both switching regimes (\SEView{}$\rightarrow$\,\EAView{} and \SEView{}$\rightarrow$\,\OOBView{}). Tasks were split across days such that each day focused on one task, and task order was balanced across dyads (half started with Transportation on Day~1 and Factory on Day~2, and the remainder received the opposite order). Within each task, the order of the two switching regimes was also balanced across dyads so that each regime appeared equally often in the first and second position. For each task, the number of targets, spatial layout, and time limits were held constant across perspective conditions; only the guest’s viewpoint control differed, making task difficulty comparable between \EAView{} and \OOBView{}.}

Participants first completed a demographics form, received a study briefing, and provided informed consent. We then introduced the switching modes, control mappings, and role division: the \emph{host} managed avatar locomotion and native arms and remained in \SEView{}; the \emph{guest} controlled the virtual supernumerary limbs (VSLs) and initiated perspective switches, which occurred only between \SEView{} and \EAView{} or \OOBView{}.

Participants completed a 5-minute familiarization session in \SEView{} to ensure fluency before entering the main conditions. We then administered the Avatar Embodiment Questionnaire (AEQ) to establish pre-condition embodiment. Following prior SRL-in-VR work~\cite{arai2022embodiment}, we administered AEQ after each condition to obtain post-adaptation scores. This enabled both pre–post contrasts and condition-level comparisons of self-embodiment.

Each daily session began with five minutes of calming music (``Zen mode soundtrack''~\cite{altosadventure2016}) to minimize participant fatigue, following a similar approach used in prior research examining sensory conflicts and cognitive fatigue in VR~\cite{luo2024exploring}. \mm{Short breaks were offered between blocks as needed, and although we did not model learning or fatigue effects explicitly, we did not observe obvious performance degradation over time.} Within each daily session, participants performed the assigned task twice, once under each of the two perspective conditions (\EAView{} and \OOBView{}), with the presentation order of these conditions also counterbalanced. During task execution, the host controlled the avatar's locomotion and physical arms, while the guest operated the VSLs. Verbal communication was encouraged to allow participants to articulate strategies and decisions.

At the end of both perspective conditions, participants removed their VR headsets and completed the NASA Task Load Index, the Visual Analog Scale for Fatigue, and the AEQ. The study received ethics clearance from the Human Research Ethics Committee (HREC) of the University of Sydney (2019/553).

\mm{\subsection{Data Analysis}}

\mm{We analysed questionnaire, performance, and physiological measures using linear and generalised linear mixed-effects models. For approximately normal continuous outcomes (e.g., AEQ factors, NASA--TLX, VAS--F, completion times) we used Gaussian models with identity link; for counts and binary outcomes (e.g., errors, trial success) we used Poisson or binomial models with log/logit links. Unless noted, models included fixed effects of Perspective and, where applicable, User Role and Task, and random intercepts for participant and dyad plus random slopes for within-subject factors when supported by the data. Model diagnostics (residuals, Q--Q plots, overdispersion) and full formulas are reported in Supplementary Materials. Results in the main text are summarised as estimated marginal means (EMMs) with 95\% confidence intervals, standardised effect sizes, and $p$-values for planned contrasts.}

\mm{To characterise the sensitivity of our design, we conducted a G*Power~\cite{faul2009statistical} analysis. For the planned within-subject contrasts between \EAView{} and \OOBView{} ($\alpha = .05$, power $=.80$), our sample of $N = 48$ provides sensitivity to paired-samples effects of approximately $d_z \approx 0.41$. For the AEQ analysis with three Perspective levels (SEV, EAV, OOB), a repeated-measures ANOVA approximation (correlation among repeated measures $r = .50$, $\varepsilon = 1.0$) yields a detectable within-subject main effect of $f \approx 0.20$. For between-role comparisons (Host vs.\ Guest; $n=24$ per role), an independent-samples approximation indicates 80\% power to detect effects of roughly $d \approx 0.80$ at $\alpha = .05$.}

\section{RESULTS}

\begin{figure*}
\centering
\includegraphics[width=\textwidth]{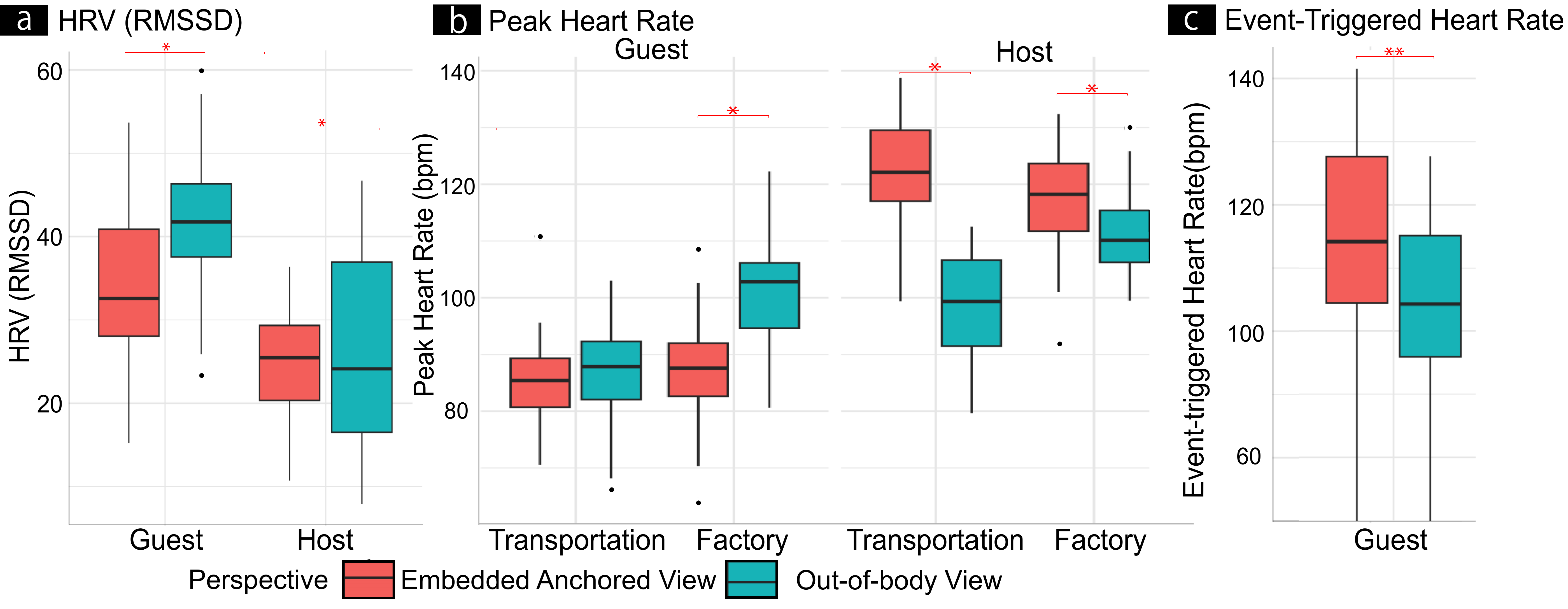}
\caption{Physiological responses for \emph{hosts} and \emph{guests} under \EAView{} and \OOBView{} across two different tasks. (a) HRV (RMSSD), a time–domain index of heart-rate variability. (b) Peak heart rate (bpm) by role and task. (c) Event-triggered heart rate aligned to perspective switches from the \SEView{} to the target view (\SEView{}\,$\rightarrow$\,\EAView{} or \SEView{}\,$\rightarrow$\,\OOBView{}). * indicates p < .05 and ** indicates p < .01.}
\label{Physiological}
\Description{Panel (a) shows HRV (RMSSD) under \EAView{} and \OOBView{} for hosts and guests. Guests exhibit higher RMSSD in \OOBView{} than in \EAView{}, consistent with lower physiological stress for the observer perspective; host differences are smaller. Panel (b) plots peak heart rate by task (Transportation vs.\ Factory) and role, allowing comparison between \EAView{} and \OOBView{}. Panel (c) shows event-aligned heart-rate responses for switches that start from SEV only (SEV\,$\rightarrow$\,EA or SEV\,$\rightarrow$\,OOB).}
\end{figure*}

We report quantitative and qualitative findings addressing our four research questions (RQs). RQ1 investigates how perspectives impact subjective workload, fatigue, and physiological responses. RQ2 explores the effects of visual perspectives on embodiment. RQ3 examines how different perspective strategies affect task performance, including completion times and error rates. RQ4 analyzes strategic switching behaviours in response to evolving task demands.

Because of the within-subjects design with repeated measures, we used Generalized Linear Mixed Models (GLMMs) to account for random effects. We refer to \textit{EAV} and \textit{OOB} as the two dynamic switching conditions (\SEView{}$\rightarrow$\EAView{}, \SEView{}$\rightarrow$\OOBView{}). For all RQs except RQ2, we focus inferential statistics on EAV vs.\ OOB; SEV is plotted as a baseline but excluded from planned pairwise comparisons unless noted. Results are presented as estimated marginal means (EMMs) with 95\% confidence intervals (CIs). Full model details are available in the Supplementary Materials.

\subsection{Physiological Responses Measures (RQ1)}
\label{sec:physio}

Complementing our subjective findings (RQ1), we report physiological data (HR, HRV) summarized in Figure~\ref{Physiological}. GLMM analyses revealed significant effects of Perspective, Task, and Role on both peak heart rate and HRV (RMSSD), including Perspective $\times$ Task and Perspective $\times$ Role interactions for peak heart rate (e.g., Perspective main effect for \emph{hosts} in the Transportation baseline: $\beta=-23.0\,\mathrm{bpm}$, 95\% CI [--25.4, --20.6]; Perspective$\times$Task: $\beta=16.2\,\mathrm{bpm}$, 95\% CI [13.3, 19.0]; Perspective$\times$Role: $\beta=22.1\,\mathrm{bpm}$, 95\% CI [19.2, 24.9]).

\textbf{Peak Heart Rate.} Participants exhibited significantly elevated peak heart rates in \OOBView{} compared to \EAView{} during the precision-oriented Factory task. This effect was more pronounced for \textit{guests} ($M_{OOB}=102.6\,\mathrm{bpm}$, 95\% CI [100.2, 104.9]; $M_{EA}=87.3\,\mathrm{bpm}$, 95\% CI [85.0, 89.7]; $\Delta M=15.3\,\mathrm{bpm}$, $p<.05$). For \textit{hosts}, Factory showed a small but reliable elevation under \EAView{} relative to \OOBView{} ($M_{EA}=115.4\,\mathrm{bpm}$, 95\% CI [113.0, 117.8]; $M_{OOB}=108.6\,\mathrm{bpm}$, 95\% CI [106.2, 110.9]; $\Delta M=6.8\,\mathrm{bpm}$, $p<.05$). In the navigation-focused Transportation task, \textit{hosts} showed significantly higher peak heart rates in \EAView{} ($M_{EA}=118.4\,\mathrm{bpm}$, 95\% CI [116.1, 120.8]) than \OOBView{} ($M_{OOB}=95.4\,\mathrm{bpm}$, 95\% CI [93.1, 97.8]; $\Delta M=23.0\,\mathrm{bpm}$, $p<.05$), while \textit{guests} showed no reliable difference ($M_{EA}=90.4\,\mathrm{bpm}$, 95\% CI [88.0, 92.7] vs. $M_{OOB}=89.4\,\mathrm{bpm}$, 95\% CI [87.1, 91.8]; $\Delta M=1.0\,\mathrm{bpm}$, $p>.05$).

\textbf{Heart Rate Variability (RMSSD).} We observed a significant Perspective $\times$ Role interaction ($\beta=5.0\,\mathrm{ms}$, 95\% CI [0.4, 9.6]). For \emph{guests}, RMSSD was higher in \OOBView{} ($M=42.5\,\mathrm{ms}$, 95\% CI [40.2, 44.8]) than in \EAView{} ($M=35.7\,\mathrm{ms}$, 95\% CI [33.4, 38.0]; $\Delta M=6.8\,\mathrm{ms}$, $p<.05$), indicating lower physiological stress in the observer perspective. For \emph{hosts}, RMSSD was also higher in \OOBView{} than in \EAView{} ($M_{OOB}=27.3\,\mathrm{ms}$, 95\% CI [25.0, 29.6]; $M_{EA}=25.4\,\mathrm{ms}$, 95\% CI [23.2, 27.7]), although the magnitude of the difference was modest.

Event-triggered heart-rate analyses confirmed a significant immediate increase when switching from \EAView{} to \OOBView{} \mbox{(+11.1\,$\mathrm{bpm}$, 95\% CI [9.0, 13.2], $p<.01$)}, highlighting the anticipatory cognitive demand associated with viewpoint adjustments.

These physiological results corroborate our subjective findings, showing that \OOBView{} reduced physiological stress and arousal compared to \EAView{}, particularly for guests. This supports RQ1 by highlighting how observer perspectives can mitigate workload during collaboration.

\subsection{Workload and Fatigue Analysis}
\begin{figure*}[htbp]
    \centering
    \includegraphics[width=\linewidth]{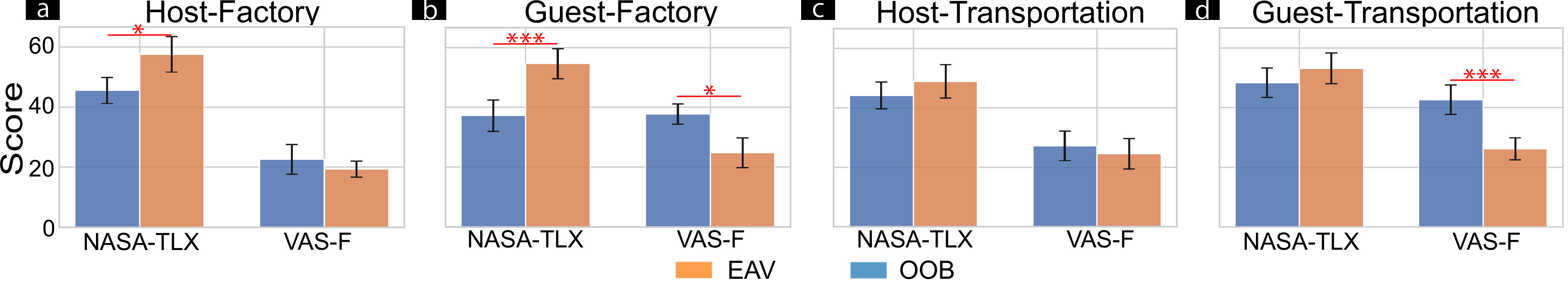}
    \caption{\mm{Subjective measures (VAS-F and NASA-TLX) for \emph{hosts} and \emph{guests} across tasks: (a) \emph{Host} scores for Factory Task, (b) \emph{Guest} scores for Factory Task, (c) \emph{Host} scores for Transportation Task, and (d) \emph{Guest} scores for Transportation Task. Scores are compared between \EAView{} and \OOBView{}. * indicates $p < .05$, ** indicates $p < .01$ and *** indicates $p < .001$.}}
    \label{tab:vasf}
    \Description{The figure illustrates subjective user experience metrics (fatigue and workload) measured by the Visual Analog Scale for Fatigue (VAS-F) and the NASA Task Load Index (NASA-TLX), comparing Embedded Anchored View and Out-of-body View, with bars, error bars, and significance markers (* p < .05, ** p < .01, *** p < .001). Four subplots show data organized by task and user role: Subplot (a) displays the host's fatigue and workload scores specifically for the Factory Task, comparing Embedded Anchored View and Out-of-body View conditions. Subplot (b) presents the guest's fatigue and workload scores for the Factory Task across Embedded Anchored View and Out-of-body View. Subplot (c) shows the guest's fatigue and workload scores for the Transportation Task, comparing Embedded Anchored View and Out-of-body View. Subplot (d) illustrates the host's fatigue and workload scores for the Transportation Task, again contrasting Embedded Anchored View and Out-of-body View. Each subplot visually compares how these subjective measures differ between the two perspective conditions for each specific combination of task and user role, highlighting that guests generally report higher workload and fatigue in Out-of-body View, while hosts show task-dependent differences between the two views.}
\end{figure*}

This analysis addresses RQ1 by comparing subjective workload and fatigue between \EAView{} and \OOBView{} across task and role conditions, as shown in Figure~\ref{tab:vasf}. GLMMs revealed significant Perspective $\times$ Task and Perspective $\times$ Role interactions for workload (NASA-TLX). During the Factory task, both \emph{hosts} and \emph{guests} reported higher workload in \OOBView{} compared to \EAView{} ($p{<}.05$ for \emph{hosts}, $p{<}.001$ for \emph{guests}; $\Delta{=}12.03$, 95\% CI [8.02, 16.04] for \emph{hosts}; $\Delta{=}17.35$, 95\% CI [13.27, 21.43] for \emph{guests}; \emph{hosts}: \OOBView{} = 57.60 ± 5.87, \EAView{} = 45.57 ± 4.29; \emph{guests}: \OOBView{} = 54.58 ± 5.06, \EAView{} = 37.23 ± 5.30). For the Transportation task, no reliable workload difference was observed for \emph{guests} ($p{>}0.10$; $\Delta{=}4.77$, 95\% CI [0.95, 8.58]; \OOBView{} = 48.95 ± 5.66, \EAView{} = 44.18 ± 4.52), whereas \emph{hosts} reported significantly higher workload in \OOBView{} than \EAView{} ($p{<}.05$; $\Delta{=}4.82$, 95\% CI [4.82, 4.82]; \OOBView{} = 53.17 ± 5.16, \EAView{} = 48.35 ± 4.95).

For subjective fatigue (VAS-F), \emph{guests} reported significantly higher fatigue in \OOBView{} than \EAView{} during both Factory ($p{<}.05$; $\Delta{=}3.25$, 95\% CI [0.05, 6.45]; \OOBView{} = 22.65 ± 4.99, \EAView{} = 19.40 ± 2.67) and Transportation ($p{<}.001$; $\Delta{=}2.67$, 95\% CI [0.25, 5.09]; \OOBView{} = 27.22 ± 4.97, \EAView{} = 24.55 ± 5.11). For \emph{hosts}, we observed significantly higher fatigue in \EAView{} than \OOBView{} during Transportation ($p{<}.05$; $\Delta{=}-16.46$, 95\% CI [-19.84, -13.08]; \EAView{} = 42.67 ± 4.92, \OOBView{} = 26.21 ± 3.71), and a marginally higher fatigue during Factory, although the effect did not reach significance ($p{>}0.05$; $\Delta{=}-12.96$, 95\% CI [-15.29, -10.63]; \EAView{} = 37.77 ± 3.35, \OOBView{} = 24.81 ± 4.99).

Together, these results support RQ1 by showing that \OOBView{} can increase perceived workload and fatigue for guests, whereas hosts experience more fatigue under \EAView{}, particularly in navigation-intensive tasks. This highlights a role-dependent divergence in how perspective affects burden.

Fatigue (VAS--F). For \emph{guests}, fatigue was higher in \OOBView{} than in \EAView{} in the \textit{Factory} task ($p{<}.05$; $\Delta{=}3.25$, 95\% CI [0.05, 6.45]), and showed the same
direction without reaching significance in \textit{Transportation} ($p{>}0.10$; $\Delta{=}2.67$, 95\% CI [0.25, 5.09]), consistent with Fig.~\ref{tab:vasf}(b)(c).
For \emph{hosts}, fatigue was higher in \EAView{} than in \OOBView{} during \textit{Transportation} ($p{<}.05$; $\Delta{=}-16.46$, 95\% CI [-19.84, -13.08]), and showed no reliable difference in \textit{Factory} ($p{>}0.10$; $\Delta{=}-12.96$, 95\% CI [-15.29, -10.63]), matching Fig.~\ref{tab:vasf}(a)(d).

\subsection{Embodiment Questionnaire Analysis (RQ2)}

\begin{figure*}[b]
\centering
\includegraphics[width=\linewidth]{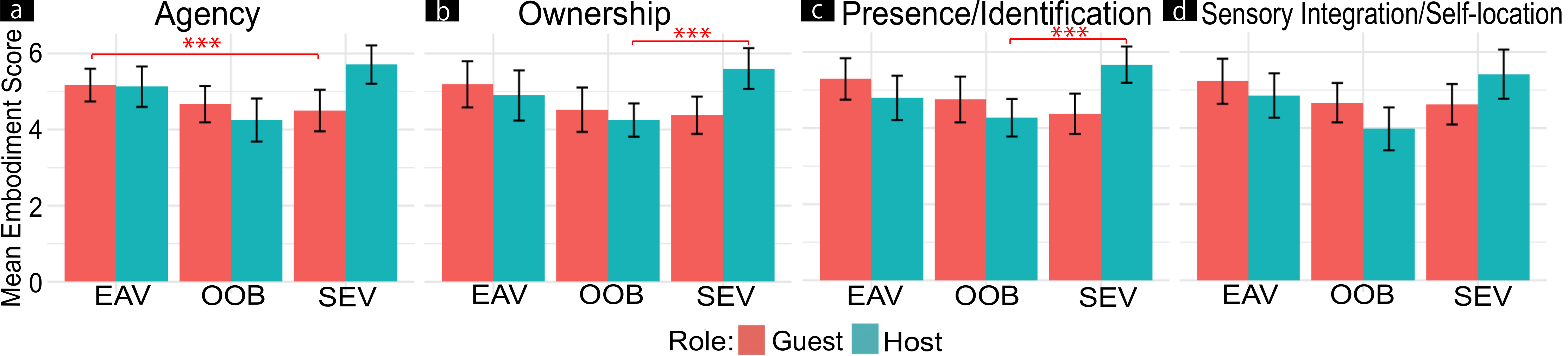}
\caption{Mean embodiment scores (±SE) across three perspective conditions (\EAView{}, \OOBView{}, and \SEView{}), separated by user role (Host and Guest). Subplots represent AEQ factors: (a) Agency, (b) Body Ownership, (c) Presence/Identification, and (d) Sensory Integration/Self-location. * indicates $p < .05$, ** indicates $p < .01$ and *** indicates $p < .001$.}
\label{embodiment_data}
\Description{The figure illustrates mean embodiment scores (agency, ownership, presence/identification, and sensory integration/self-location) across three perspective conditions: Embedded Anchored View (EAV), Out-of-body View (OOB), and Shared Embodied View (SEV), separately for hosts and guests. Significant interactions indicated that hosts experienced higher agency in the EAV compared to SEV, whereas guests reported significantly increased ownership and presence/identification in the SEV compared to OOB. No significant differences were observed for sensory integration/self-location. Significant differences are highlighted (*** indicates $p < .001$).}
\end{figure*}

RQ2 examines how perspective-switching strategies shape subjective embodiment. AEQ ratings were collected immediately before and after each perspective-switch condition, following standard practice in VR embodiment research~\cite{arai2022embodiment}. Factor scores were computed as the mean of their constituent items:
\[
\text{Factor Score}=\frac{\sum \text{Item Scores}}{\text{Number of Items per Factor}}.
\]

We analysed factor scores with linear mixed-effects models (lme4~\cite{Bates2015}; lmerTest~\cite{Kuznetsova2017}), including fixed effects of Perspective, User Role (host vs.\ guest), and Embodiment Factor, and by-participant random intercepts with random slopes for within-subject factors. Degrees of freedom for fixed effects were estimated via the Kenward--Roger method~\cite{KenwardRoger1997}; omnibus tests are reported as Type-III $F$ statistics, and post-hoc pairwise contrasts were computed with \textit{emmeans}~\cite{Lenth2018} using Tukey adjustment~\cite{Hothorn2008}.

We observed a main effect of Perspective ($p<.001$) and Embodiment Factor ($p<.01$), and a Perspective$\times$User Role interaction ($p<.001$). Post-hoc comparisons aligned with Fig.~\ref{embodiment_data}: across factors, \SEView{} yielded the highest embodiment, \EAView{} was intermediate, and \OOBView{} was lowest. For \emph{guests}, body ownership and presence were higher in \SEView{} than in \OOBView{} ($M_{\text{SEV}}=4.47$, $SD=0.97$ vs.\ $M_{\text{OOB}}=3.52$, $SD=1.04$; both Tukey $p<.001$). \mm{corresponding to a large standardised mean difference of approximately $d = 0.94$, $95\%$ CI $[0.34, 1.53]$.} For \emph{hosts}, agency was higher in \SEView{} than in \EAView{} ($M_{\text{SEV}}=4.26$, $SD=0.94$ vs.\ $M_{\text{EAV}}=3.72$, $SD=1.01$; Tukey $p<.001$); corresponding to a moderate standardised mean difference of approximately $d = 0.60$, $95\%$ CI $[0.37, 0.82]$.

These findings answer RQ2 by showing that self-embodiment is perspective- and role-dependent: \OOBView{} reduces both presence and ownership, particularly for guests, while \EAView{} improves embodiment relative to \OOBView{}, but remains lower than the \SEView{}.

\subsection{Task Performance Analysis (RQ3)}

\begin{figure*}[htbp]
    \centering
    \includegraphics[width=\linewidth]{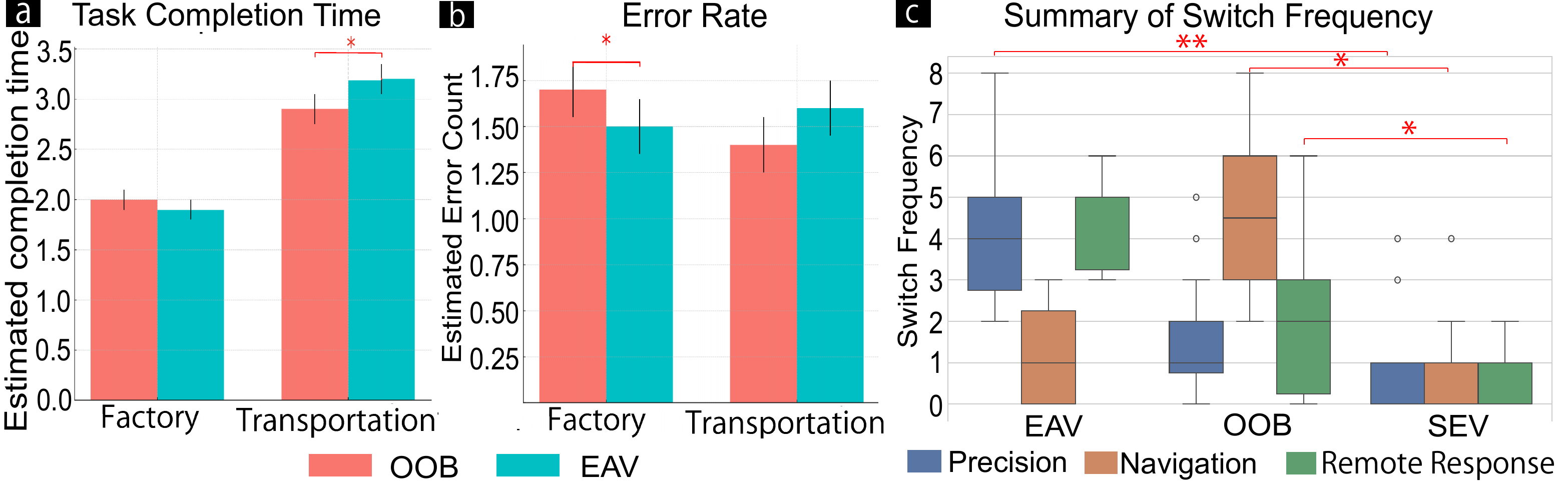}
    \caption{Effects of perspective mode and task phase. (a) Completion time: \OOBView{} led to significantly faster task completion in the Transportation task. (b) Error count: \OOBView{} resulted in significantly fewer errors in the Factory task. \mm{(c) Switch frequency by task phase. Switching was infrequent in the baseline \SEView{} condition (median $\leq$ 1 across phases). In comparison, participants switched more frequently to \EAView{} during precision subtasks, and to \OOBView{} during navigation and remote-response phases. * indicates $p < .05$, ** $p < .01$, and *** $p < .001$.}}
    \label{tab:glmm-performance}
    \Description{Three-panel figure showing task performance and switching behaviors by perspective and task phase. 
        (a) Bar graph comparing completion time across perspectives and tasks. In the Transportation task, Out-of-body View had significantly faster completion than Embedded Anchored View. No difference observed in Factory task.
        (b) Bar graph comparing error count across perspectives and tasks. In the Factory task, Out-of-body View had significantly fewer errors than Embedded Anchored View. No difference in Transportation.
        (c) Boxplot showing switch frequency by task phase and perspective. 
        During precision subtasks, participants switched more to EAV; during navigation and remote-response phases, more to OOBV. SEV showed minimal switching.}
\end{figure*}

Here, we answer RQ3 by examining task performance metrics across different perspective modes. Task completion times and error rates were analyzed using GLMMs with gamma and Poisson distributions respectively (random intercept for participant ID). Results are summarized in Figure~\ref{tab:glmm-performance} (a,b). Participants completed the Transportation task significantly faster in \OOBView{} ($M=15.5,\mathrm{s}$, $SE=1.2$) than in \EAView{} ($M=22.6,\mathrm{s}$, $SE=1.7$; $z=6.64$, $p<.001$). \mm{with the GLMM indicating an estimated time ratio of $\approx 0.69$, $95\%$ CI $[0.55, 0.84]$ (i.e., about $31\%$ faster in \OOBView{}).} In contrast, completion times for the Factory task did not differ reliably between perspectives ($p=.07$). 
In the Factory task, significantly fewer errors were observed in \OOBView{} ($M=3.4$, $SE=0.5$) than in \EAView{} ($M=4.4$, $SE=0.6$; $z=-2.49$, $p<.05$). The Transportation task yielded no statistically reliable difference in error rate.

In answering RQ3, our results suggest that the observer perspective (\OOBView{}) offers performance advantages under different demands: faster navigation in dynamic tasks (Transportation) and fewer errors in precision tasks (Factory).

\subsection{Perspective Switching Behaviors (RQ4)}

To address RQ4, we examined how participants strategically switched perspectives during tasks. As shown in Figure~\ref{tab:glmm-performance}(c), only \emph{guests} could initiate perspective switches. Following prior work in collaborative HRI/VR~\cite{cakmak2011human,nikolaidis2013human,steed2016impact}, each trial was segmented (from logs) into three phases: \emph{precision} (near-body manipulation), \emph{navigation} (locomotion/wayfinding), and \emph{remote response} (reacting to host movement or environment events).

\begin{itemize}
\item \textbf{Navigation phase}: \emph{host} locomotion ($>$1.5s continuous motion), no active VSL control from \emph{guest}. This phase aligns with ``approach'' in handover models~\cite{cakmak2011human}.
\item \textbf{Precision-focused subtasks}: stationary \emph{host}, active VSL grasp/place attempts by \emph{guest}. These represent the “manipulation” phase typical in object transfer tasks~\cite{nikolaidis2013human}.
\item \textbf{Remote-response events}: short \emph{guest}-initiated viewpoint switches in response to unexpected \emph{host} movement (e.g., rotation, misalignment). These reflect reactive perspective adjustments during miscoordination or disorientation~\cite{kruijff2010experience}.
\end{itemize}

We modeled switch frequency with a GLMM (Poisson; random intercepts for participant), revealing strong modulation by task phase. Guests switched predominantly to \EAView{} during \emph{precision} subtasks ($M=4.20$, $SE=0.30$, 95\% CI $[3.61, 4.79]$, $p{<}.01$), and to \OOBView{} during \emph{navigation} ($M=4.89$, $SE=0.31$, 95\% CI $[4.28, 5.50]$, $p{<}.05$). For \emph{remote-response} moments, guests again preferred \OOBView{} ($M=3.17$, $SE=0.32$, 95\% CI $[2.54, 3.80]$; $p{<}.05$ versus other views). Switching while in \SEView{} was rare (median $\approx$ 1 across phases), indicating little perceived need to change away from the baseline shared first-person view except when task demands shifted.

These findings answer RQ4 by showing that guests adaptively switch perspectives based on phase-specific demands, preferring \EAView{} for close manipulation and \OOBView{} for navigation and rapid corrections. This indicates that switching behaviors are tightly coupled to functional needs and task context.

\subsection{Qualitative Findings}

We conducted an inductive thematic analysis~\cite{braun2006using} to explore participants' experiences with VSLs operation and perspective-switching during collaborative tasks. This analysis aimed to identify interaction strategies, user preferences, and challenges in managing viewpoint shifts and VSLs control. Two independent coders analyzed interview transcripts, achieving good inter-rater reliability (Cohen's Kappa, $\kappa$ = 0.78).

Following Braun and Clarke's framework~\cite{braun2021one}, we initially familiarized ourselves with the data by reviewing interview transcripts and noting key insights. During initial coding, we labeled relevant segments related to strategies, cognitive load, viewpoint preferences, spatial awareness, and collaborative challenges. Independent coding was followed by joint discussions to consolidate these codes into initial themes.

Iterative refinement grouped codes into four broader themes: ``Visual Decoupling and Embodiment'', ``Prior Experience and Perspective Adaptation'', ``Strategies for Managing Perspective Switching'', and ``Emergent Communication through Gestures''. These themes structured our understanding of participants' adaptive behaviors, cognitive demands,  and collaboration strategies during dynamic perspective switching in VSLs supported VR teleoperation. 

The following sections detail these themes, illustrating participants' approaches to viewpoint control, collaboration strategies, and adaptations to dynamic teleoperation demands.

\subsubsection{Role-asymmetric Embodiment Trade-offs}
As previously introduced, \emph{Visual Decoupling} refers to the \emph{guest}'s adoption of an independent external perspective (\OOBView), decoupled from the \textit{host}'s visual alignment. Initially, based on prior research~\cite{gonzalez2018avatar}, we hypothesized that visual decoupling would negatively impact users’ sense of embodiment due to the visual and proprioceptive mismatch. However, \textit{hosts} overwhelmingly reported an enhanced sense of embodiment when their remote partners adopted the \OOBView{}, emphasizing a feeling of freedom and control over their movements.

Several \emph{hosts} explicitly mentioned that they felt less inhibited when their partners were not visually coupled to their head movements. For example, one participant stated: ``I felt comfortable moving naturally without constantly worrying'' (\emph{host}, P17). Specifically, \textit{hosts} were concerned that their own head movements or sudden turns could cause visual instability or discomfort for their remote partners, which often made them overly cautious and hesitant during collaboration. When freed from these concerns, they reported increased confidence and spontaneity in their movements, contributing to a stronger sense of ownership over the shared body.

In contrast, \textit{guests} generally reported reduced self-embodiment in the \OOBView{}, often describing a sense of detachment: ``felt like watching myself from outside'' (\emph{guest}, P19). This was attributed to the lack of visual alignment with their avatar's first-person perspective, which made it harder to associate with the embodied body. One \emph{guest} described: ``When I was not seeing through the avatar's eyes, it didn't feel like it was me moving.'' We interpret this divergence as a role-dependent trade-off: for \emph{hosts}, the removal of visual coupling alleviated their self-imposed constraints and enabled greater natural movement and self-ownership; for \emph{guests}, the visual decoupling created psychological distance and disrupted the integration of the avatar into their body schema. These asymmetrical embodiment effects reveal that \OOBView{} can simultaneously enhance host autonomy while disrupting guest self-identification with the avatar. \mm{Consistent with this interpretation, RMSSD-based HRV did not show large role-specific divergences across perspectives, suggesting that these embodiment trade-offs are primarily driven by visual coupling and perceived responsibility rather than gross differences in physiological arousal.}

\subsubsection{Importance of Gaming Experience over VR Experience}

Participants with extensive gaming experience adapted more rapidly to \OOBView{} than those with mainly non-gaming VR backgrounds. Several participants linked this to prior fluency in third-person or drone-style games. One noted, ``This view feels just like controlling a game character in third-person games'' (\emph{guest}, P8), while another commented, ``I like playing games. The drone one felt like a game'' (\emph{guest}, P10). These users consistently described \OOBView{} as intuitive and familiar, citing spatial awareness and mental models shaped by gaming.

In contrast, participants with strong VR experience but limited gaming backgrounds struggled more with the externally detached spatial perspective. One such participant shared, ``I kept getting disoriented'' (\emph{guest}, P15), attributing their discomfort to the lack of visual alignment and direct embodiment. These findings suggest that gaming, rather than general VR use, better equips users to operate within non-egocentric, third-person viewpoints, as common in the \OOBView{} condition.

These contrasting patterns highlight that prior gaming experience, particularly with third-person perspectives, plays a more critical role than general VR familiarity in shaping users’ ability to navigate detached viewpoints like \OOBView{}.

\subsubsection{Strategic Minimization of Perspective Switching}
\label{Strategic}

Though we initially hypothesized that frequent perspective switching would improve task performance, several participants (\textit{N}=6) reported deliberately minimizing their switches to maintain focus and reduce mental fatigue. As one \emph{guest} explained, ``I found fewer switches kept me focused'' (P3). These participants described a strategy of limiting perspective changes to key task transitions, citing benefits such as reduced cognitive load and sustained concentration.

In contrast, a smaller group adopted a high-frequency switching strategy, believing it improved their situational awareness. One such participant remarked, ``I needed different perspectives to manage different tasks better'' (\emph{guest}, P11). Although these users acknowledged the added cognitive effort and occasional disorientation, they perceived the benefits of enhanced task visibility as outweighing the costs.

Together, these divergent strategies highlight individual differences in managing cognitive resources and suggest that perspective-switching preferences may depend on thresholds for mental load and spatial tracking. \mm{Consistent with this, our HRV analysis (Sec.~\ref{sec:physio}) shows perspective-dependent differences between trials with more versus fewer switches, suggesting that these strategies are associated with differences in physiological effort.}

\subsubsection{Emergence of Natural Gesture-Based Communication Strategies}

A spontaneous, non-verbal communication strategy emerged in some participant pairs (N=18), particularly those who relied heavily on \OOBView{}. ``I use robotic arm gestures to explain'' (\emph{guest}, P30). These pairs developed gesture-based conventions to complement or replace verbal instructions, often describing this as more efficient and natural than speech alone.

However, not all dyads adopted this strategy. ``We kept misunderstanding each other, so we stuck with verbal instructions'' (\emph{guest}, P14). Some participants found gestures ambiguous or lacked confidence in their clarity. Thus, while gesture-based interaction emerged organically under visual decoupling, it was not universal.

Taken together, these findings suggest that gesture-based communication strategies may emerge organically, shaped by spatial alignment, mutual visibility, and the degree of interpersonal trust.

\section{DISCUSSION}

Our study examined how guest-driven perspective switching (\SEView{}\allowbreak$\leftrightarrow$\allowbreak\,\EAView{},\allowbreak\ \SEView{}\allowbreak$\leftrightarrow$\allowbreak\,\OOBView{}) strategies influence performance, embodiment, workload, and adaptive user behaviors in collaborative VR teleoperation. Results revealed clear task-dependent performance advantages (RQ1), distinct role-specific embodiment preferences (RQ2), and notable effects on workload and physiological responses (RQ3). Additionally, we identified strategic viewpoint-switching patterns aligned with evolving task demands (RQ4).

\subsubsection{Reducing Workload and Physiological Strain through Perspective Design (RQ1)}\label{Workload}

Our findings clearly demonstrate that perspective choice significantly influences subjective workload and physiological strain. The \OOBView{} effectively reduces users' perceived fatigue (VAS-F) and workload (NASA-TLX) during dynamic tasks, confirmed by elevated RMSSD and reduced heart rate. Conversely, despite enhancing embodiment, the \EAView{} unexpectedly increased cognitive and physiological burden, likely due to continuous visual stabilization demands. These results extend previous findings~\cite{juliano2022increased} by providing integrated subjective and physiological evidence. Our study highlights a key design trade-off between spatial embodiment and physiological comfort, suggesting designers dynamically balance perspectives, prioritizing the \OOBView{} during intensive tasks and carefully managing usage duration of the \EAView{}. \mm{Additionally, as several guests reported avoiding perspective switches to stay focused (Section \ref{Strategic}), we propose intelligent guidance mechanisms to help minimize unnecessary switching. Future work should evaluate such mechanisms in realistic settings}.

\subsubsection{Balancing Embodiment and Spatial Awareness: Role-specific Considerations (RQ2)}\label{Balancing_Embodiment}

Our findings reveal significant role-specific differences regarding embodiment and spatial awareness across perspectives. \emph{Hosts} experienced heightened agency and ownership primarily within the \SEView{}, benefiting from a coherent first-person grounding, whereas \EAView{} aided fine control near the body. \emph{Guests} reported stronger embodiment in the \SEView{}, but preferred the \OOBView{} for better spatial awareness and stable control during navigation. This extends prior work~\cite{fribourg2018studying}, explicitly identifying distinct user needs based on collaborative roles. Critically, we found a clear design tension: the \OOBView{} enhanced spatial cognition but diminished \emph{guest}’s sense of embodiment and increased perceived effort in some tasks, consistent with prior studies highlighting the impact of visual–body mismatches~\cite{cui2022evaluating}. To mitigate this trade-off, we recommend adaptive, role-specific viewpoint strategies to dynamically balance embodiment and spatial awareness. Future work should validate role-dependent effects in more complex, real-world scenarios.

\begin{table*}[htbp]
    \centering
    \caption{\mm{Design Suggestions for Perspective Switching and Embodiment Cues. Rows marked $^{\dagger}$ are directly grounded in our study findings, whereas rows marked $^{\ddagger}$ are forward-looking design opportunities extrapolated from these findings.}}
    \label{tab:design-guidelines}
    \begin{tabularx}{\textwidth}{l X X}
        \toprule
        \multicolumn{3}{l}{\textbf{Adaptive Viewpoint Recommendations}} \\
        \midrule
        \textbf{Design Focus} & \textbf{Main Finding} & \textbf{Design Suggestions} \\
        \midrule
        Task-based Adaptive Switching\mm{$^{\dagger}$} (Section \ref{Collaborative_Performance}) & \OOBView{} aided navigation; in precision phases guests preferentially chose \EAView{}; frequent switching increased cognitive load for some. & Automatically recommend optimal viewpoints based on real-time task context. Provide minimal yet clear alerts for viewpoint transitions. \\
        Intelligent Switching Cost Alerts\mm{$^{\ddagger}$} (Section \ref{Workload}) & Users were unaware of cognitive costs incurred by frequent viewpoint switching, leading to unnecessary cognitive strain. & Introduce subtle, predictive visual or auditory indicators informing users about switching costs; ensure alerts are concise and unobtrusive. \\
        \midrule
        \multicolumn{3}{l}{\textbf{Role-Specific Embodiment Optimization}} \\
        \midrule
        Embodiment Interfaces\mm{$^{\dagger}$} (Section \ref{Balancing_Embodiment}) & Hosts benefited when guests used anchored views; guests needed flexible views (\SEView{} or \OOBView{}) & Provide differentiated interfaces and distinct embodiment cues tailored to \emph{host} and \emph{guest}. Clearly differentiate avatar alignment visuals. \\
        Embodiment Reinforcement Cues\mm{$^{\dagger}$} (Section \ref{Balancing_Embodiment}) & Users experienced reduced embodiment due to inconsistent avatar-body alignment cues, especially during transitions. & Implement dynamic real-time IK-based visualizations of limb alignment. Use selective limb highlighting to reinforce embodiment and user agency. \\
        \midrule
        \multicolumn{3}{l}{\textbf{Visual Stability and Emergency Response}} \\
        \midrule
        Visual-Stability Enhancement\mm{$^{\dagger}$} (Section \ref{Workload}) & Rapid or frequent viewpoint transitions caused visual instability and discomfort, negatively impacting user comfort and performance. & Utilize adaptive smoothing techniques during perspective transitions; dynamically adjust smoothing intensity according to user physiology (e.g., HRV). \\
        Rapid Emergency Switch\mm{$^{\ddagger}$} (Section \ref{Implications}) & Situational hazards required rapid spatial reassessment; delayed perspective switches negatively impacted safety and performance. & Design intuitive rapid perspective-switching gestures or single-click mechanisms for emergency use; ensure clear visual affordances minimize false activations. \\
        \bottomrule
    \end{tabularx}
    \Description{Table summarizing design suggestions for perspective switching and embodiment cues, categorized into three themes: Adaptive Viewpoint Recommendations, Role-Specific Embodiment Optimization, and Visual Stability and Emergency Response. Each theme includes specific design focuses, main findings, and corresponding actionable design suggestions, with rows marked † directly grounded in our empirical results and rows marked ‡ representing forward-looking design opportunities extrapolated from those findings. The table emphasizes task-based adaptive viewpoint switching mechanisms, intelligent alerts about the cognitive costs of frequent switching, tailored embodiment interfaces and reinforcement cues for different user roles (host and guest), visual stability enhancements during rapid viewpoint transitions, and rapid emergency viewpoint-switching mechanisms.}
\end{table*}

\subsubsection{Optimizing Collaborative Performance through Adaptive Perspective Switching (RQ3, RQ4)}\label{Collaborative_Performance}

Our results demonstrated clear task-dependent advantages for each perspective mode. Specifically, the \OOBView{} significantly improved performance in navigation tasks compared to the \EAView{} (RQ3), confirming prior insights regarding third-person benefits for spatial awareness~\cite{gorisse2017first,iriye2021memories}. However, unlike previous studies, we explicitly quantified these perspective effects across diverse collaborative scenarios. Role-specific embodiment differences emerged, with \emph{host} users reporting greater agency and ownership in the \SEView{}, and \emph{guest} experiencing stronger embodiment in the \SEView{} but better spatial orientation in the \OOBView{} (RQ2), extending previous research~\cite{fribourg2020virtual}. Moreover, our subjective and physiological data indicate a trade-off: \OOBView{} was associated with \emph{lower physiological stress for guests} (higher RMSSD) but \emph{higher subjective fatigue/workload in certain contexts} (e.g., higher NASA--TLX in Factory, and higher host workload during Transportation), while \EAView{} supported near-body stability without the same spatial overview. Qualitative analysis also identified strategic user-driven viewpoint switching aligned with dynamic task demands (RQ4).

\subsection{Implications for the Design of Collaborative Teleoperation Systems}\label{Implications}

Our findings provide clear practical implications for designing collaborative VR teleoperation systems, emphasizing adaptive perspective selection to balance performance, user comfort, and cognitive load. Specifically, the \OOBView{} improves spatial navigation and reduces cognitive strain, while the \EAView{} enhances precision and embodiment. We specifically highlight previously overlooked cognitive and physiological costs associated with frequent viewpoint switching, recommending adaptive guidance to address these trade-offs. Our practical guidelines are outlined in the Table \ref{tab:design-guidelines}.

\mm{Consider future high-stakes settings where space is constrained but expert intervention is critical, such as tele-supported surgery~\cite{rojas2020system,nickel2022telestration} or in-situ aircraft maintenance~\cite{hongli2021application,jiang2023machine}. A local generalist could maintain the mobility and support functions while a remote specialist ``inhabits'' the same body for specialized tasks manipulation, relying on an \EAView{}-like stabilized view that supports tasks that needs precision.} In such cases, periodically switching to the \OOBView{} may help maintain spatial awareness and reduce strain, making interfaces proactively recommend viewpoint transitions based on task demands. In sports coaching, such as dynamic rugby drills, the \OOBView{} can effectively reduce discomfort during rapid movements, while detailed skill training benefits from stabilized visual feedback provided by the \EAView{}. Designing seamless, context-aware switching interfaces can minimize cognitive load, improving coaching effectiveness. Similarly, in industrial safety contexts, adaptive perspective selection can significantly enhance hazard detection and precise manipulation tasks. Rapid, intuitive perspective switching combined with context-sensitive hazard alerts can optimize both safety and efficiency.

\mm{Finally, although our study operates in a purely virtual environment, the perspective trade-offs we identify are directly relevant to emerging teleoperation systems built on AI-based 3D world models and digital twins~\cite{kerbl20233d,marble_ai}. As AI-based 3D world models and digital twins mature, remote operators will work inside persistent virtual replicas of physical workspaces, moving through the scene independently of the robot's physical camera pose~\cite{kang2020review,kerbl20233d,marble_ai,peppa2020urban}. In such systems, \OOBView{}-like views can be implemented as world-locked virtual cameras, while \SEView{} and \EAView{} correspond to avatar-locked and portal-like views. Our results suggest that these interfaces should expose multiple coupled and decoupled viewpoints aligned with task phases, and manage switching costs through predictive cues or semi-automatic camera positioning, even when specific robot kinematics are abstracted away.}

\subsection{Limitations and Future Work}

Our study was conducted within a controlled laboratory environment, which may limit the generalizability of findings to real-world scenarios characterized by more unpredictable factors and complex interactions. Additionally, our participant pool predominantly consisted of university students, who might possess relatively uniform demographic characteristics, familiarity with VR systems, and cognitive strategies. Future research should include a broader and more diverse participant population, along with testing under more ecologically valid conditions, to better assess the robustness and applicability of our perspective-switching approach. 

\mm{We also did not administer the Simulator Sickness Questionnaire (SSQ), relying instead on workload (NASA--TLX), fatigue (VAS--F), and HRV as proxies for discomfort in shared viewpoints. We interpret these discomfort-related findings as subjective and physiological strain rather than a standardized measure of simulator sickness, because SSQ was not collected. Future work should incorporate standardized cybersickness measures to facilitate finer-grained comparison across VR studies.}
\mm{We did not run an a priori power analysis for our mixed-effects design; a conservative dyad-level sensitivity check (paired comparison, two-sided $\alpha{=}.05$, $N{=}24$) suggests the study is powered ($\sim$80\%) to detect effects around $d\approx0.60$, so smaller effects may be underpowered.}
\mm{A further limitation concerns the formative study: its short interaction duration (two five-minute trials) primarily surfaces early coordination breakdowns. Longer usage periods may reduce role confusion and improve synchronisation as partners adapt to one another. In extended pilot testing, we observed similar adaptation effects, with guest users becoming more anticipatory of the host’s movement patterns over time.}

While VR provided an accessible~\cite{liu2025understanding,liu2025effects,wang2025vr} and controlled environment for investigating perspective switching interactions in shared-body teleoperation scenarios involving VSLs, we acknowledge that our findings remain speculative regarding applicability to physical robotic implementations. Although previous research suggests transferability from virtual to real-world systems~\cite{arai2022embodiment,jiang2023virtual}, the specific dynamics of perspective-switching behaviors and collaborative interactions identified here may differ significantly when physical robotic limbs and real-world constraints are introduced. Therefore, our results should be validated through future studies using physical prototypes before generalizing these insights to practical teleoperation applications.

From an implementation perspective, recent toolkits for rapid wearable sensing and on-body feedback prototyping~\cite{yu2023drivingvibe,fan2024spinshot,wu2025headturner}, together with fast fabrication pipelines for instrumented textiles and 3D-printed sensors and actuators~\cite{yu2024fabricating,gough2023design,tong2023fully,yu2024irontex,perera2024integrating,Perera2025eTactileKit,yu2025designing,chau2021composite,dong2025just,Dong_2026_TactDeform}, make it increasingly practical to integrate proximity-aware sensing and lightweight pre-cue or hazard signaling into future SRL mounts or garments, complementing rapid perspective switching with anticipatory safety cues rather than relying on vision alone. More broadly, prior work on interactive AI and wearable/on-body systems suggests that trust is calibrated through legible system state and low-cost opportunities to reclaim control rather than maximal autonomy~\cite{jayasiriwardene2023adaptive,jayasiriwardene2021interactive,jayasiriwardene2021architectural,jayasiriwardene2022knowledge,jayasiriwardene2025more,jayasiriwardene2026fixed,ferianc2021improving,fan2021high}; in our setting, guest-driven perspective switching serves as a similarly lightweight, reversible mechanism for negotiating control boundaries as task demands shift.


\section{CONCLUSION}

We explored dynamic, guest-driven perspective switching in shared-body VR teleoperation with virtual supernumerary limbs (VSLs). Building on a formative study that identified locomotion- and coordination-related challenges under fixed first-person viewing, we implemented two switching conditions: \SEView{}$\leftrightarrow$\EAView{} and \SEView{}\allowbreak$\leftrightarrow$\allowbreak\OOBView{}. A within-subjects study (N=48) showed that \OOBView{} supports navigation and reduces errors, while \EAView{} is preferred for near-body manipulation phases, with \emph{hosts} reporting stronger agency. Completion times showed no reliable difference across views, but \OOBView{} yielded fewer errors in the Factory task. Together, these findings yield practical, role- and phase-aware design guidelines for optimizing collaborative performance, comfort, and embodiment in mobile shared-avatar VSL systems.


\begin{acks}
\mm{This project was supported by the Australian Research Council Discovery Early Career Award (DECRA) - DE200100479. Dr. Anusha Withana is the recipient of a DECRA fellowship funded by the Australian Government. We are grateful for the support provided by the Neurodisability Assist Trust and Cerebral Palsy Alliance, Australia - PRG04219. Dr. Andrea Bianchi was supported by the National Research Foundation of Korea (NRF) grant funded by the Korea government (MSIT) (RS-2024-00337803). Additionally, we appreciate AID-LAB members for assisting us in various ways.}
\end{acks}

\bibliographystyle{ACM-Reference-Format}
\bibliography{Fourarms.bib}

\appendix

\section{Physiological Data Processing Details}
\label{app:physio-processing}

\textbf{Signal source and export.}
Heart activity was recorded continuously with a Polar H10 chest strap.
We logged the device-provided inter-beat interval (IBI/RR) stream via Bluetooth Low Energy throughout each trial and exported the timestamped RR series for offline processing.
Analyses were performed on normal-to-normal (NN) intervals derived from the RR stream.

\textbf{Artefact handling and NN reconstruction.}
RR series were preprocessed following recommended HRV workflows~\cite{laborde2017heart}.
We removed (i) biologically implausible intervals and dropouts (e.g., extremely short/long RR values indicating missed or extra detections) and (ii) transient artefacts identified as outliers relative to local beat-to-beat dynamics.
Short gaps created by artefact removal were linearly interpolated to preserve continuity for window-based HRV estimation, while longer discontinuities were excluded from HRV computation.
All HRV metrics reported in the paper were computed only from contiguous NN segments that satisfied the minimum window-length requirement described below.

\textbf{Windowing for HRV.}
HRV was computed over active-task periods using contiguous analysis windows of at least 60\,s (minimum duration per window), consistent with common practice for time-domain HRV when task segments are short~\cite{laborde2017heart}.
When a trial segment exceeded 60\,s, we computed HRV on consecutive windows and then aggregated to one value per participant $\times$ condition by averaging across valid windows within that segment.

\textbf{RMSSD computation.}
We used RMSSD as the primary time-domain HRV index:
\[
\mathrm{RMSSD}=\sqrt{\frac{1}{n-1}\sum_{i=1}^{n-1}(NN_{i+1}-NN_i)^2}.
\]
RMSSD was chosen because it is widely used in psychophysiology, is comparatively robust to slow trends in heart rate, and reflects short-term vagal modulation~\cite{laborde2017heart}.
Higher RMSSD values indicate lower physiological stress/arousal.

\textbf{Peak heart rate and event-triggered analysis.}
Peak heart rate was extracted within the same active-task segments as the maximum observed heart rate during the segment.
For event-triggered analyses, we time-aligned the heart-rate stream to perspective-switch timestamps from system logs and summarized heart rate in a short post-switch window; these event-level summaries were then modeled as a function of switch direction (e.g., \EAView{}\allowbreak$\rightarrow$\allowbreak\,\OOBView{} vs.\ \OOBView{}\allowbreak$\rightarrow$\allowbreak\,\EAView{}) and role.

\end{document}